\documentclass{article}

\usepackage{arxiv}

\usepackage[utf8]{inputenc} 
\usepackage[T1]{fontenc}    
\usepackage{hyperref}       
\usepackage{url}            
\usepackage{booktabs}       
\usepackage{amsfonts}       
\usepackage{nicefrac}       
\usepackage{microtype}      
\usepackage{lipsum}		
\usepackage{graphicx}
\usepackage{natbib}
\usepackage{doi}
\usepackage{amsmath}

\title{Robust Full Waveform Inversion with deep Hessian deblurring}

\author{ \href{https://orcid.org/0009-0005-5239-3955}{\includegraphics[scale=0.06]{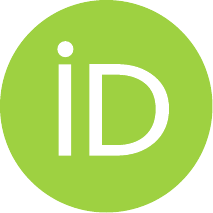}\hspace{1mm}Mustafa Alfarhan} \\
	KAUST\\
	Thuwal, Kingdom of Saudi Arabia\\
	\texttt{mustafa.alfarhan@kaust.edu.sa} \\
	\And
	{Matteo Ravasi} \\
	KAUST\\
	Thuwal, Kingdom of Saudi Arabia\\
	\texttt{matteo.ravasi@kaust.edu.sa} \\
        \And
	{Fuqiang Chen} \\
	KAUST\\
	Thuwal, Kingdom of Saudi Arabia\\
	\texttt{fuqiang.chen@kaust.edu.sa} \\
        \And
	{Tariq Alkhalifah} \\
	KAUST\\
	Thuwal, Kingdom of Saudi Arabia\\
	\texttt{tariq.alkhalifah@kaust.edu.sa} \\
}

\renewcommand{\shorttitle}{A deep inverse Hessian for FWI}

\hypersetup{
pdftitle={A deep inverse Hessian for FWI},
pdfsubject={FWI},
pdfauthor={Mustafa Alfarhan},
pdfkeywords={Full waveform inversion, Imaging; Born modeling, Deep learning, Neural networks},
}

\begin{document}
\maketitle

\begin{abstract}
Full Waveform Inversion (FWI) is a technique widely used in geophysics to obtain high-resolution subsurface velocity models from waveform seismic data. Due to its large computation cost, most flavors of FWI rely only on the computation of the gradient of the loss function to estimate the update direction, therefore ignoring the contribution of the Hessian. Depending on the level of computational resources one can afford, an approximate of the inverse of the Hessian can be calculated and used to speed up the convergence of FWI towards the global (or a plausible local) minimum. In this work, we propose to use an approximate Hessian computed from a linearization of the wave-equation as commonly done in Least-Squares Migration (LSM). More precisely, we rely on the link between a migrated image and a doubly migrated image (i.e., an image obtained by demigration-migration of the migrated image) to estimate the inverse of the Hessian. However, instead of using non-stationary compact filters to link the two images and approximate the Hessian, we propose to use a deep neural network to directly learn the mapping between the FWI gradient (output) and its Hessian (blurred) counterpart (input). By doing so, the network learns to act as an approximate inverse Hessian: as such, when the trained network is applied to the FWI gradient, an enhanced update direction is obtained, which is shown to be beneficial for the convergence of FWI. The weights of the trained (deblurring) network are then transferred to the next FWI iteration to expedite convergence. We demonstrate the effectiveness of the proposed approach on two synthetic datasets and a field dataset.
\end{abstract}


\section{Introduction}
Seismic data, commonly acquired at the surface of the Earth, contain useful information about the Earth's structure. Seismic waves travel in the subsurface, interact with it, and some of them eventually return to the surface to be recorded by an array of receivers. However, inferring the subsurface structure and its medium properties from seismic data is a challenging task. This difficulty arises mostly from the fact that the subsurface is accessed from only one side, and therefore some areas are not covered well by the recorded waves, or in other words, they are poorly illuminated. Moreover, multiply scattered waves (i.e., free-surface and internal multiples) interfere with primary arrivals, which present a direct linear link with the subsurface properties of interest. As a consequence, geophysical research over the years has been focused on developing methods that can separate different parts of the seismic data and optimally utilize them to image the subsurface. Full Waveform Inversion (FWI) was introduced by~\cite{lailly1983seismic,tarantola1984inversion} as an alternative approach that aims to utilize the entire seismic waveform data at once to construct a high-resolution subsurface model. More precisely, FWI casts the problem of transforming seismic records into a subsurface model as an optimization problem, where a measure of the difference between modeled and observed data is minimized. Given an initial subsurface model, obtained, for example, through traveltime tomography~\cite{bording1987applications} or  migration velocity analysis~\cite{symes2008migration}, synthetic seismic data are calculated with a wave-equation based forward modeling engine. The data misfit between modeled and observed data is then utilized to update the model parameters and the process is repeated until convergence.

As a consequence of its pursuit in utilizing the entire seismic data content at once, FWI is known to be a highly non-linear inverse problem. Whether or not FWI converges to the right model depends on the initial model: more specifically, if the travel-time difference between the modeled and observed data is more than half a cycle, we will converge to a local minimum model, a problem known as cycle-skipping. To mitigate cycle-skipping, FWI is usually performed in a multi-scale fashion~\cite{bunks1995multiscale,miller1996multiscale,fichtner2013multiscale,alkhalifah2016full,zhang2019seismic} where the inversion is applied to the low-frequency components of the data first. Only after the kinematic part of the subsurface model is correctly inverted for to ensure that we are within the half-cycle difference for the higher frequencies that the high-frequency components of the data are incorporated to add high-resolution features to the inverted model. Although there exist both local~\cite{lailly1983seismic,tarantola1984inversion} and global~\cite{jin1994nonlinear} approaches to solving FWI, the former is usually preferred: each model update involves calculating the first and (possibly) second derivatives of the data-misfit with respect to the current model, which are called the gradient and the Hessian, respectively. Whilst the gradient provides the direction of the model update, the inverse of the Hessian alters the direction slightly in compliance with the curvature information and introduces a parameter-dependent scaling to the gradient. From a physical standpoint, the inverse of the Hessian accounts for losses in amplitude due to geometrical spreading and transmission losses occurring in the wave equation, as well as it provides a trade-off between different model parameters in multi-parameter FWI~\cite{alkhalifah2016full}. However, calculating the inverse of the Hessian for large subsurface models is computationally expensive.

Traditionally, gradient-based methods have been popularized in FWI applications as they only rely on the gradient for the model update calculation. Among gradient-based methods, the most popular ones are the steepest-descent method, whose update is given by a scaled version of the negative of the gradient, and the non-linear conjugate-gradient method, whose update is a combination of the gradient at the current iteration and the previous update. On the other hand, second-order methods such as the Newton method rely on the Hessian in addition to the gradient to define the model update. A special case, when second-order derivative terms in the Hessian (which account for second-order scattering) are neglected, is represented by the Gauss-Newton (GN) method ~\cite{pratt1998gauss}. Different strategies have emerged over the years to approximate the GN Hessian. For example, the gradient can be preconditioned by the diagonal elements of the Hessian or the pseudo-Hessian as proposed by~\cite{shin2001improved}. On the other hand, the model and gradient from previous iterations can be used to iteratively update the inverse of the Hessian, as done in the family of quasi-Newton methods such as the limited-memory BFGS algorithm~\cite{liu1989limited}. Another alternative to estimating the Hessian is the family of truncated-Newton methods, which do not need the Hessian explicitly but rather compute the Hessian-vector product~\cite{metivier2014full,altheyab2013time}. An equivalently accurate approach to approximate the GN Hessian, and is the inspiration to the approach developed here, is a least-squares optimization of the gradient~\cite{doi:10.1190/geo2014-0365.1}. In general, the inclusion of the approximate inverse of the Hessian provides faster convergence than gradient-based methods.

Seismic imaging is another area where approximating the Hessian is of great importance as the resolution of the resulting seismic images can be enhanced and uneven illumination resulting from imperfect acquisition geometries can be mitigated. More specifically, Least-Squares Migration (LSM) is a technique that aims to mitigate the degrading effects of the Hessian. In the image domain, the Hessian effect can be considered a form of blurring operator. Therefore, LSM can be formulated as an image deblurring problem where the inverse of the Hessian is the deblurring operator. Using a migrated and doubly migrated image, the inverse of the Hessian can be approximated with non-stationary matching filters as done in~\cite{guitton2004amplitude}. Similarly,~\cite{aoki2009fast} use deblurring filters to accelerate LSM in a regularized or preconditioning fashion. Alternatively, as initially proposed by~\cite{valenciano2008imaging}, one can rely on point spread functions (PSFs) to construct the deblurring filters. In recent years, various deep learning methods have been proposed for image deblurring (which are equivalently applicable to the problem of approximating the action of the inverse of the Hessian matrix). \cite{liu2022deep} suggested to employ a convolutional neural network to learn the deconvolution operator of the PSFs in least-squares reverse time migration (LSRTM). \cite{kumar2022deep} uses a neural network to learn the inverse of the Hessian using the original migration image and a remigrated image. Finally, \cite{peng2023} suggest to use a loop-unrolled gradient descent method to include the contribution of the known PSF in the learning process. In all studies, it was concluded that using a network to approximate the action of the inverse of the Hessian is faster than other state-of-the-art methods and can more stably deconvolve seismic images.

In this work, we propose to embed a similar approach to that of~\cite{kumar2022deep} within FWI. More specifically, Born modeling and its adjoint are used to create our Hessian blurred gradient. In this case, the Hessian blurred gradient is approximately related to the FWI gradient through the inverse of the Hessian, whose mapping is learned using a deep neural network. After training, the network is applied to the gradient to obtain a Hessian corrected update that we demonstrate can speed up FWI convergence. We apply our approach to two synthetic data to showcase its benefits, and to the Volve field data to validate its applicability under field conditions.


\section{Theoretical Framework}
\subsection{Full waveform inversion}
The constant-density acoustic wave equation is given by
\begin{equation}\label{eq:eq1}
\frac{1}{m^2(\mathbf{x})}\frac{\partial^2 u(\mathbf{x},t)}{\partial t^2} - \nabla^2 u(\mathbf{x},t) = f(\mathbf{x},t)
\end{equation}
\noindent where $m(\mathbf{x})$ is the medium velocity, $u(\mathbf{x},t)$ is the wavefield, $f(\mathbf{x},t)$ is the source function, all defined in space $\mathbf{x}$ and time $t$, and $\nabla^2$ is the Laplacian operator. Starting from an initial velocity model, the wave equation is solved using a forward modeling scheme such as the finite difference method to obtain the simulated data, $\mathbf{d}_s$. Then, the simulated data are compared to the observed data, $\mathbf{d}_o$, using a misfit function such as the $L_2$ norm:
\begin{equation}\label{eq:eq2}
J(\mathbf{m}) = \frac{1}{2}\Delta \mathbf{d}^T \Delta \mathbf{d}
\end{equation}
\noindent where $\Delta \mathbf{d} = \mathbf{d}_o - \mathbf{d}_s$. Note that in the above equation we have expressed both datasets and the model in vector form. From now on, the use of regular or bold symbols will discriminate whether we are considering a scalar or a vector for our equations.

Using the generalized Taylor series expansion around a background model $\mathbf{m}_o$, the misfit function can be approximated by
%
\begin{equation}\label{eq:eq3}
J(\mathbf{m})  \approx J(\mathbf{m}_o) +  \left(\frac{\partial J(\mathbf{m}_o)}{\partial \mathbf{m}}\right)^T(\mathbf{m} - \mathbf{m}_o)
 + \frac{1}{2}(\mathbf{m} - \mathbf{m}_o)^T \frac{\partial^2 J(\mathbf{m}_o)}{\partial \mathbf{m}^2}(\mathbf{m} - \mathbf{m}_o)
\end{equation}
Taking the derivative of equation (\ref{eq:eq3}) with respect to $\boldsymbol\delta \mathbf{m} = \mathbf{m} - \mathbf{m}_o$ and setting it to zero yields
\begin{equation}\label{eq:eq4}
\boldsymbol\delta \mathbf{m} = - \left(\frac{\partial^2 J(\mathbf{m}_o)}{\partial \mathbf{m}^2}\right)^{-1} \frac{\partial J(\mathbf{m}_o)}{\partial \mathbf{m}}
\end{equation}
\noindent where $\boldsymbol\delta \mathbf{m}$ is the model update, $\partial J(\mathbf{m}_o)/ \partial \mathbf{m}$ is the gradient, and $\left(\partial^2 J(\mathbf{m}_o) / \partial \mathbf{m}^2\right)^{-1}$ is the inverse of the Hessian. FWI is an optimization process where the current model $\mathbf{m}$ is iteratively updated by generating the data $\mathbf{d}_s$ and comparing them against the observed data to compute the model update $\boldsymbol\delta \mathbf{m}$ accordingly until the data residual is sufficiently small. 
\subsection{Born modeling}
In Born modeling, the model $m(\mathbf{x})$ is commonly split into two parts: background model and perturbation. The background model is often smooth, while the perturbation may lead to scattering in the wavefield. Therefore, the velocity $m(\mathbf{x})$ can be written as a combination of a background velocity $m_o(\mathbf{x})$ and a perturbation $\delta m(\mathbf{x})$
\begin{equation}\label{eq:eq5}
    m(\mathbf{x}) = m_o(\mathbf{x}) + \delta m((\mathbf{x})
\end{equation}
In addition, we can approximate $1/{m^2(\mathbf{x})}$ using Taylor series as
\begin{equation}\label{eq:eq6}
    \frac{1}{m^2(\mathbf{x})} \approx \frac{1}{m_o^2(\mathbf{x})} -2 \frac{\delta m(\mathbf{x})}{m_o^3(\mathbf{x})}
\end{equation}
Similarly, the seismic wavefield $u(\mathbf{x},t)$ can be split into a background wavefield $u_o(\mathbf{x},t)$, satisfying the wave equation for $m_o(\mathbf{x})$ and a scattering wavefiled $\delta u(\mathbf{x},t)$ such that 
\begin{equation}\label{eq:eq7}
    u(\mathbf{x},t) = u_o(\mathbf{x},t) +  \delta u(\mathbf{x},t)
\end{equation}
Substituting equations (\ref{eq:eq6}) and (\ref{eq:eq7}) into equation (\ref{eq:eq1}) yields
\begin{multline}\label{eq:eq8}
    \frac{1}{m_o^2(\mathbf{x})}\frac{\partial^2 u_o(\mathbf{x},t)}{\partial t^2} - \nabla u_o(\mathbf{x},t) + \frac{1}{m_o^2(\mathbf{x})} \frac{\partial^2 \delta u(\mathbf{x},t)}{\partial t^2} \\ 
    - \nabla \delta u(\mathbf{x},t) -2\frac{\delta m(\mathbf{x})}{m_o^3(\mathbf{x})} \frac{\partial^2 (u_o(\mathbf{x},t) + \delta u(\mathbf{x},t))}{\partial t^2} = f(\mathbf{x},t)
\end{multline}
%
\noindent which can be split into the background and the scattering parts, respectively:
\begin{equation}\label{eq:eq9}
    \frac{1}{m_o^2(\mathbf{x})}\frac{\partial^2 u_o(\mathbf{x},t)}{\partial t^2} - \nabla u_o(\mathbf{x},t)= f(\mathbf{x},t)
\end{equation}
%
\begin{equation}\label{eq:eq10}
   \frac{1}{m_o^2(\mathbf{x})} \frac{\partial^2 \delta u(\mathbf{x},t)}{\partial t^2} - \nabla \delta u(\mathbf{x},t) \\
   = 2\frac{\delta m(\mathbf{x})}{m_o^3(\mathbf{x})} \frac{\partial^2 (u_o(\mathbf{x},t) + \delta u(\mathbf{x},t))}{\partial t^2}
\end{equation}
Equation (\ref{eq:eq10}) is implicit since both the left- and right-hand sides depends on the scattering wavefield. To make it explicit, we use the Born approximation, which assumes that the scattered wavefield is small compared to the background wavefield,
\begin{equation}\label{eq:eq11}
    \frac{1}{m_o^2(\mathbf{x})} \frac{\partial^2 \delta u(\mathbf{x},t)}{\partial t^2} - \nabla \delta u(\mathbf{x},t) = 2\frac{\delta m(\mathbf{x})}{m_o^3(\mathbf{x})} \frac{\partial^2 u_o(\mathbf{x},t)}{\partial t^2}
\end{equation}

Consequently, Born modeling can be performed by first modeling the background wavefield $u_o(\mathbf{x},t)$ in the background model $m_o(\mathbf{x})$ with source function $f(\mathbf{x},t)$ using equation (\ref{eq:eq9}). This wavefield is then used to ignite a secondary source in equation (\ref{eq:eq10}); in other words, a second step of modeling is therefore performed in the background model to compute the scattered wavefield using this secondary source. Note that, in practice, equations (\ref{eq:eq9}) and (\ref{eq:eq10}) are jointly solved so that there is no need to store the entire background wavefield from the first modeling to be injected in the second one. 

In compact notation, we can define the linear Born modeling as 
\begin{equation}\label{eq:eq12}
    \boldsymbol\delta \mathbf{u} = \mathbf{L} \boldsymbol\delta \mathbf{m}
\end{equation}
with $\boldsymbol\delta \mathbf{m} = \boldsymbol\delta \mathbf{m}/\mathbf{m}_o$. Finally, an additional sampling operator $\mathbf{P}$ is used to extract the  scattered wavefield at the location of the receivers and produce the surface seismic data:
\begin{equation}\label{eq:eq12b}
    \mathbf{d} = \mathbf{PL} \boldsymbol\delta \mathbf{m}  = \tilde{\mathbf{L}}\boldsymbol\delta \mathbf{m}
\end{equation}
where, for simplicity, we define  $\tilde{\mathbf{L}}$ from here onwards as the Born modeling operator.
\subsection{Approximating the inverse of the Hessian for FWI with deep learning}
The full model update for FWI as shown in equation (\ref{eq:eq4}) requires the calculation of the gradient and the inverse of the Hessian. However, the calculation of the inverse of the Hessian is highly expensive: this has led to the development of methods to estimate an approximate version of such an inverse with affordable computational cost. In this work, we propose a method to approximate the Hessian that relies on deep learning. The full Hessian is composed of two terms: a single-scattering and a double-scattering terms. In the Gauss-Newton method, the second term is neglected and only the first term is used. Multiplying the gradient by the inverse of the Hessian can be viewed as a deblurring operation that corrects the gradient for the effects of the Earth such as geometric spreading. We present an approach for approximating the inverse of the Hessian for FWI in the following sub-section.
\subsubsection{The Born modeling and its adjoint approach}
One way to approximate the Hessian is via a migration/demigration approach, or, after linearizing FWI, Born modeling and its adjoint. The gradient of FWI can be summarized as follows
\begin{equation}\label{eq:eq13}
\mathbf{g} = \tilde{\mathbf{L}}^T \Delta \mathbf{d}
\end{equation}
\noindent where $\tilde{\mathbf{L}}^T$ is the adjoint of the Born modeling operator, and $\Delta \mathbf{d}$ is the data residual. Moreover, considering the true Earth perturbation is $\boldsymbol\delta \mathbf{m}$, $\Delta \mathbf{d} = \tilde{\mathbf{L}}\boldsymbol\delta \mathbf{m}$, equation (\ref{eq:eq13}) can be equivalently written as
\begin{equation}\label{eq:eq14}
\mathbf{g} = \tilde{\mathbf{L}}^T \tilde{\mathbf{L}}\boldsymbol\delta \mathbf{m}
\end{equation}
\noindent which reveals that the gradient $\mathbf{g}$ and the true perturbation $\boldsymbol\delta \mathbf{m}$ are linked via a pair of Born modeling and its adjoint (demigration/migration) operations. A blurred gradient (referred in this work as doubly migrated image) can be obtained from the gradient by applying once again Born modeling and its adjoint
\begin{equation}\label{eq:eq15}
 \boldsymbol\delta \mathbf{m}_1 = \tilde{\mathbf{L}}^T \tilde{\mathbf{L}}\mathbf{g}
\end{equation}

Given availability of both $\mathbf{g}$ and $\boldsymbol\delta \mathbf{m}_1$, one can approximate $(\tilde{\mathbf{L}}^T \tilde{\mathbf{L}})^{-1}$ (the inverse of the Hessian) and apply it to $\mathbf{m}$ to obtain an estimate of $\boldsymbol\delta \mathbf{m}$, which is the objective of LSM. In our approach, $\mathbf{m}$ is the FWI gradient, $\mathbf{g}$. In other words, by estimating and applying the inverse of the Hessian to the FWI gradient, we aim to retrieve an approximation of the scattering potential as used in Born modeling.

\begin{figure*} 
\centering
\includegraphics[scale=0.6]{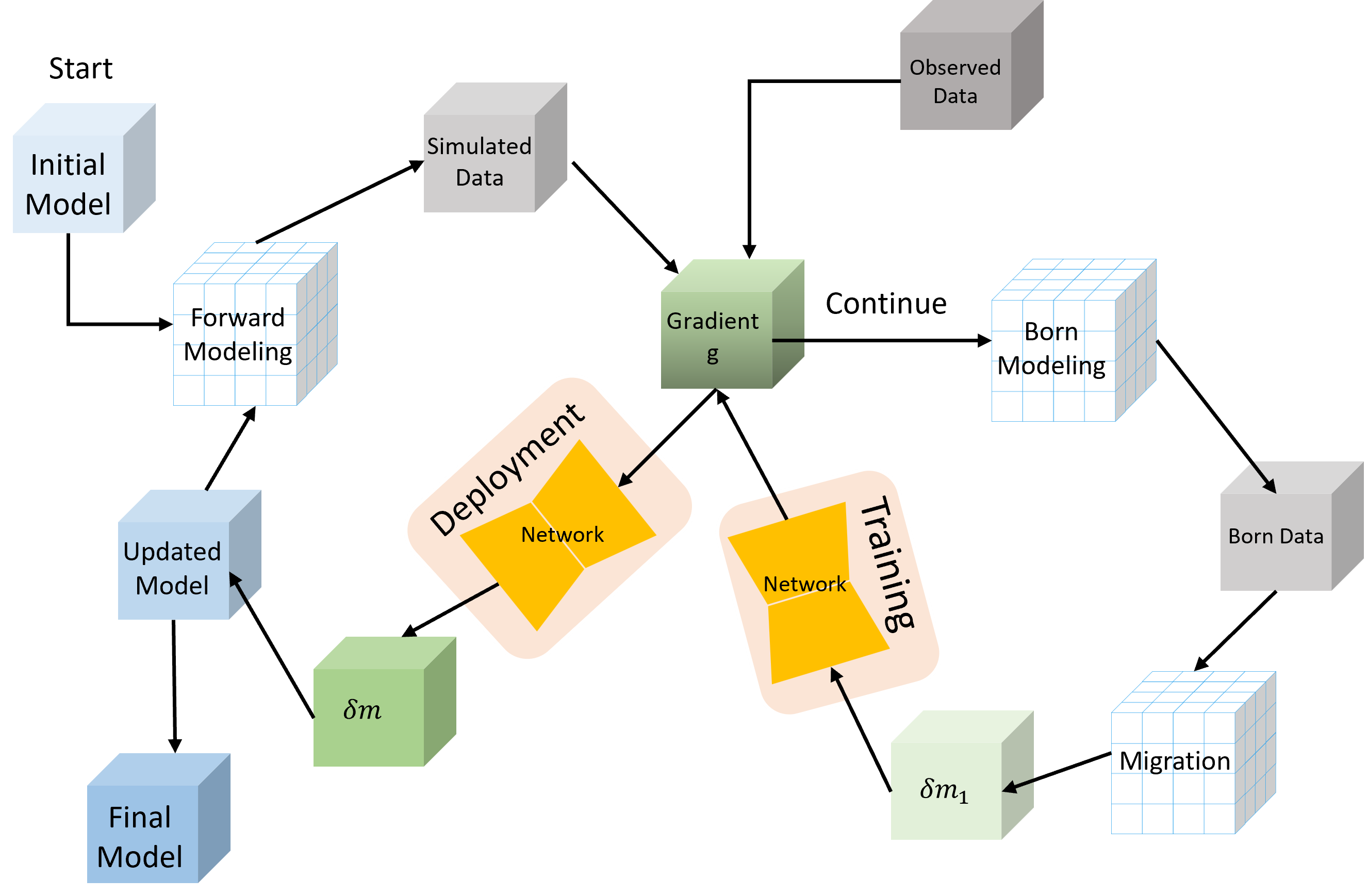} 
\caption{Proposed method to approximate the inverse of the Hessian using deep learning.} 
\label{fig_algo} 
\end{figure*}

In this work, we suggest to train a neural network to learn the mapping from $\boldsymbol\delta \mathbf{m}_1$ to $\mathbf{g}$, which ultimately represents the action of the inverse of the Hessian on a vector. Also, since the FWI update is related to the gradient through the Hessian (see equation (\ref{eq:eq4})), the trained network can be applied to the gradient to obtain an improved FWI update;  this can ultimately lead to faster convergence of the FWI process. The same process is repeated at every iteration, except for the fact that the neural network is not trained from scratch: instead, we leverage transfer learning and start from the weights learned in the previous iteration. The proposed method is depicted in Figure~\ref{fig_algo}.
\subsection{Neural-network based inverse Hessian approximation}
In this work, we suggest to train a neural network to approximate the inverse of the Hessian in FWI. Given that learning the action of the inverse of the Hessian can be framed as a domain translation process (from the doubly migrated to the  migrated image), our network presents a UNet-like architecture~\cite{ronneberger2015u}. More specifically, the network architecture is composed of four blocks in both the encoder and decoder parts. The input is passed to an input block which consists of a double convolution layer which itself contains two sequences of operations. Each sequence involves a 2D convolution with a kernel size of $3\times 3$, stride of 1, and padding of 1, followed by a Leaky ReLU activation function with a negative slope of 0.2. The first convolutional layer transforms the input channel to 128 channels, and the second maintains these 128 channels. There are four downsampling blocks in the encoder, each consisting of an average pooling operation with a $3\times 3$ kernel and a stride of 2 followed by a double convolution block (similar to the input block). These blocks progressively increase the number of channels while reducing the spatial dimensions of the feature maps. The channel sizes double at each block, going from 128 to 256, then to 512, 1024, and finally 2048. Corresponding to the downsampling blocks, there are four upsampling blocks in the decoder. Each block starts with a transposed 2D convolution which halves the number of channels, and is followed by a double convolution block (similar structure as the encoder). These blocks incrementally reduce the number of channels while increasing the spatial dimensions of the feature maps, effectively recovering the spatial resolution that was lost during downsampling. The final part of the network is an output convolution layer. It consists of a convolution layer with a kernel size of $1\times 1$, used to transform the 128 channels from the last upsampling block to the desired number of output channels (in this case, 1 channel). The network architecture is summarized in Table~\ref{tab:unet} in Appendix A. The Adam optimizer is used for optimizing the network's parameters with a cosine annealing scheduler for the learning rate. The MSE loss function is used to measure the misfit between the network's output and the FWI gradient. The training process can be carried out using patches or the full image at once. Whilst both approaches have similar performance, we chose the latter because it makes training faster.


\section{Numerical Examples}
In this section, we test the approach on two synthetic examples and one field data used to validate the proposed methodology. 

\subsection{Marmousi model}
\begin{figure}[!ht]
\centering
\includegraphics[width=\linewidth]{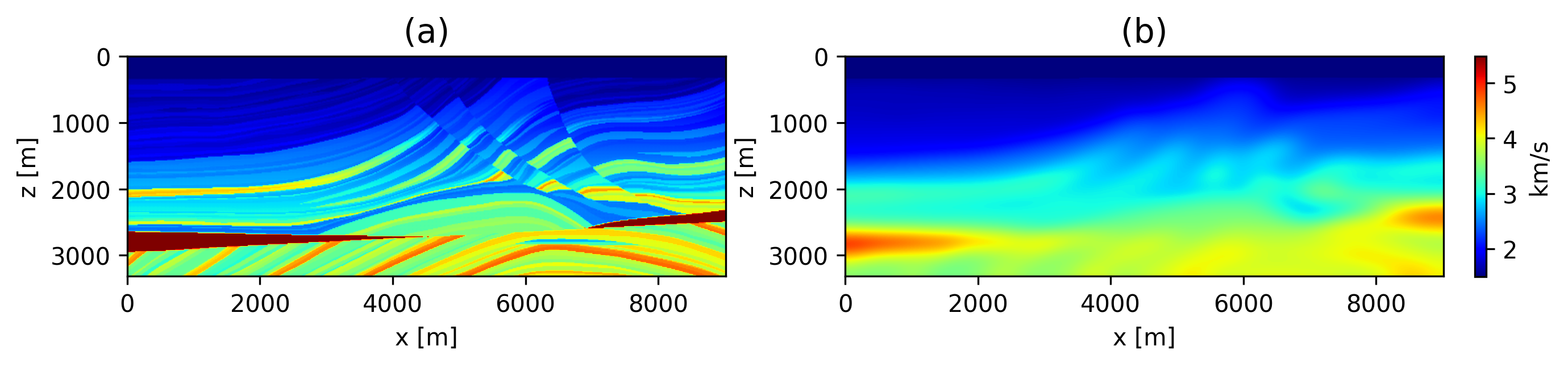} 
\caption{a) the true Marmousi model and b) its smoothed version.} 
\label{fig_marm} 
\end{figure}

\begin{figure}
\centering
\includegraphics[width=0.5\linewidth]{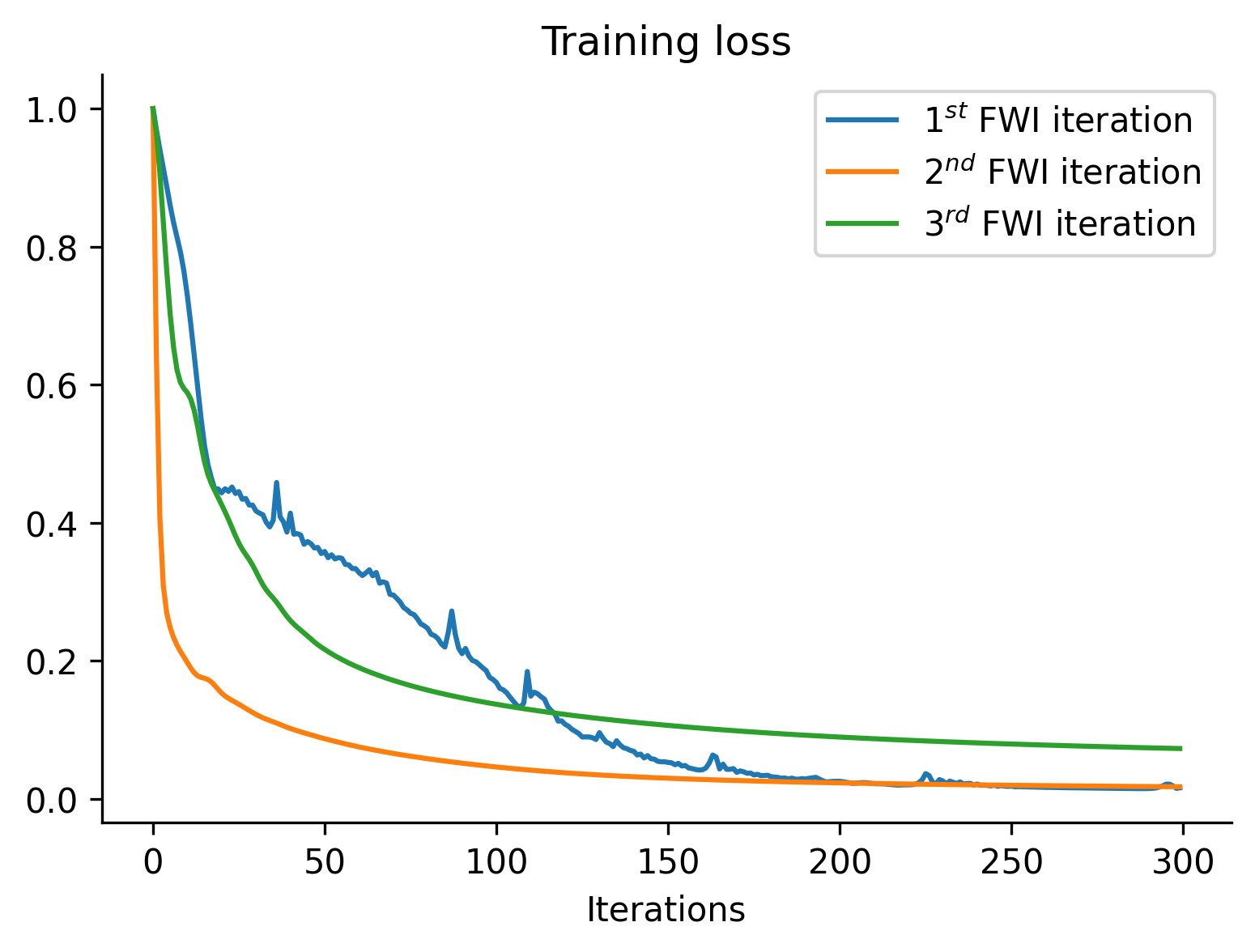} 
\caption{The normalized training loss for the first three iterations of FWI.} 
\label{fig_marm_train_loss} 
\end{figure}

\begin{figure}
\centering
\includegraphics[width=\linewidth]{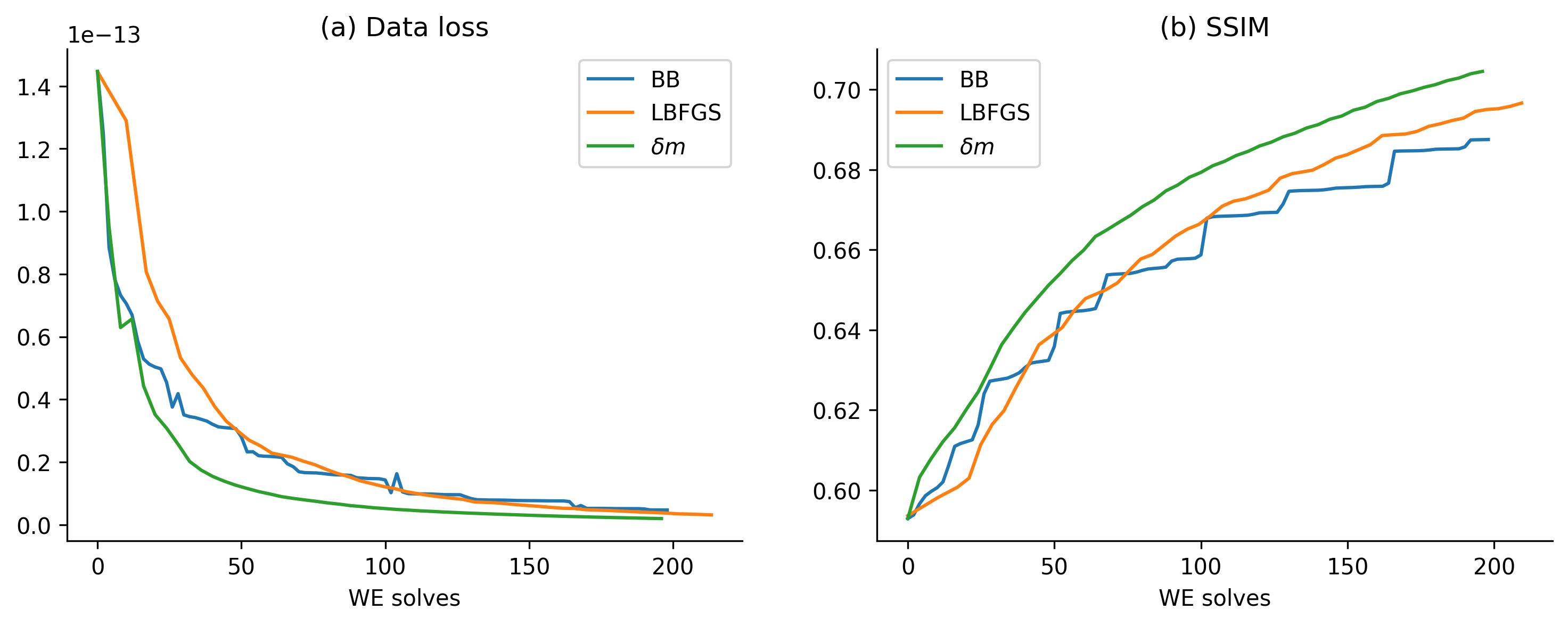} 
\caption{Data loss and SSIM index for FWI on Marmousi using the BB method, L-BFGS, and $\delta m$.} 
\label{fig_loss_marm} 
\end{figure}

\begin{figure}
\centering
\includegraphics[width=\linewidth]{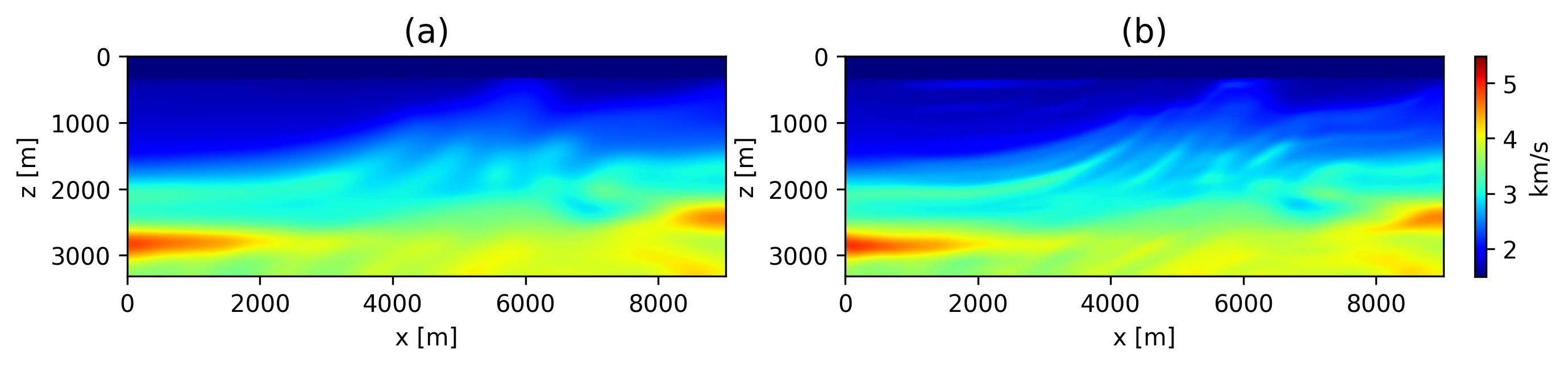} 
\caption{First model update using a) the BB method, and b) $\delta m$.} 
\label{fig_marm_iter1} 
\end{figure}

\begin{figure*}
\centering
\includegraphics[width=\linewidth]{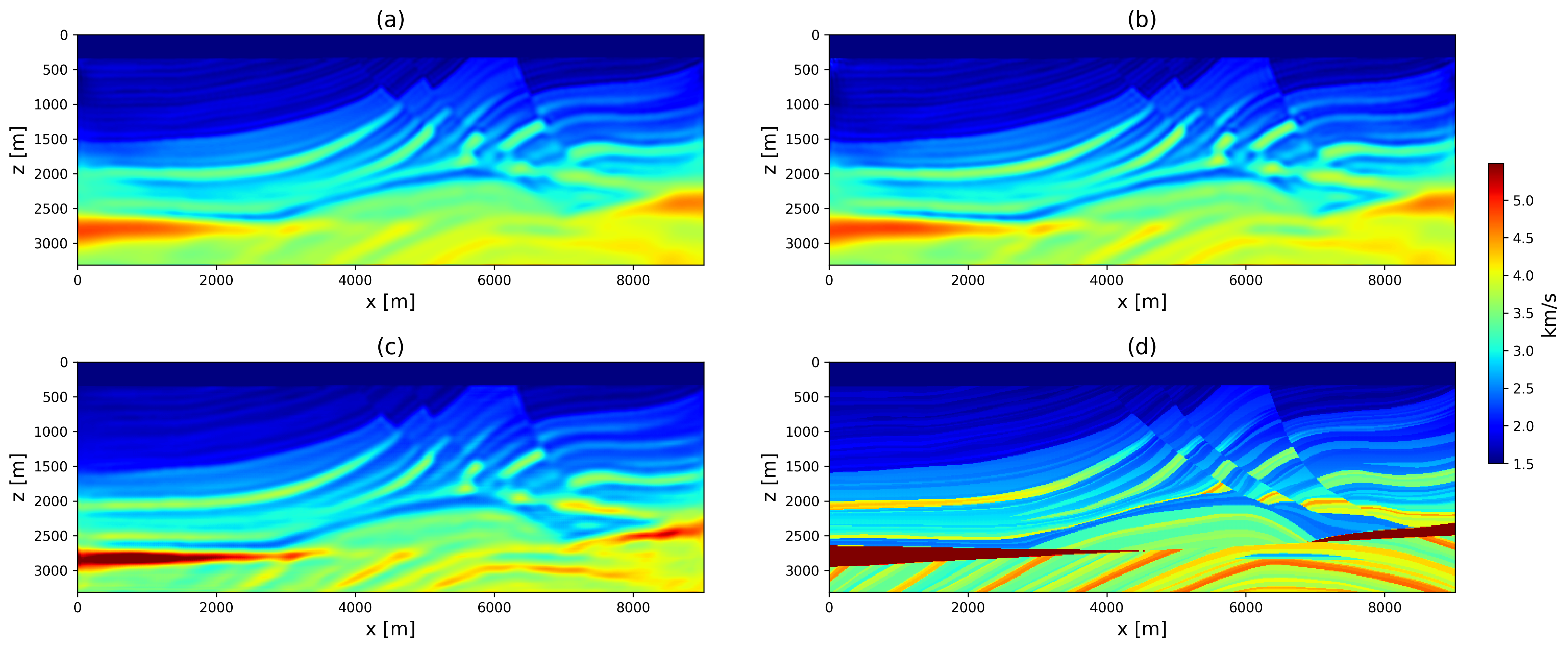} 
\caption{Final updated models: a) using the BB method, b) using L-BFGS, c) using $\delta m$}, and d) the true model for comparison. 
\label{fig_marm_iter_last} 
\end{figure*}

Due to its intricate geological composition, the synthetic Marmousi model~\cite{brougois1990marmousi} serves as a benchmark for assessing algorithms in seismic imaging and inversion. The model used in this example spans $601$ grid points horizontally and $221$ grid points vertically, with a grid size of $15\ m$ in both directions. Figure~\ref{fig_marm} illustrates both the Marmousi model and its smoothed counterpart, utilized as the starting model for FWI. The observed data is modeled using a variable-velocity, constant-density acoustic finite-difference modeling engine, using 30 sources and 300 receivers equally spaced at zero depth. A Ricker wavelet of 5 Hz peak frequency is used for the source signature.\\
\begin{figure}[!ht]
\centering
\includegraphics[width=\linewidth]{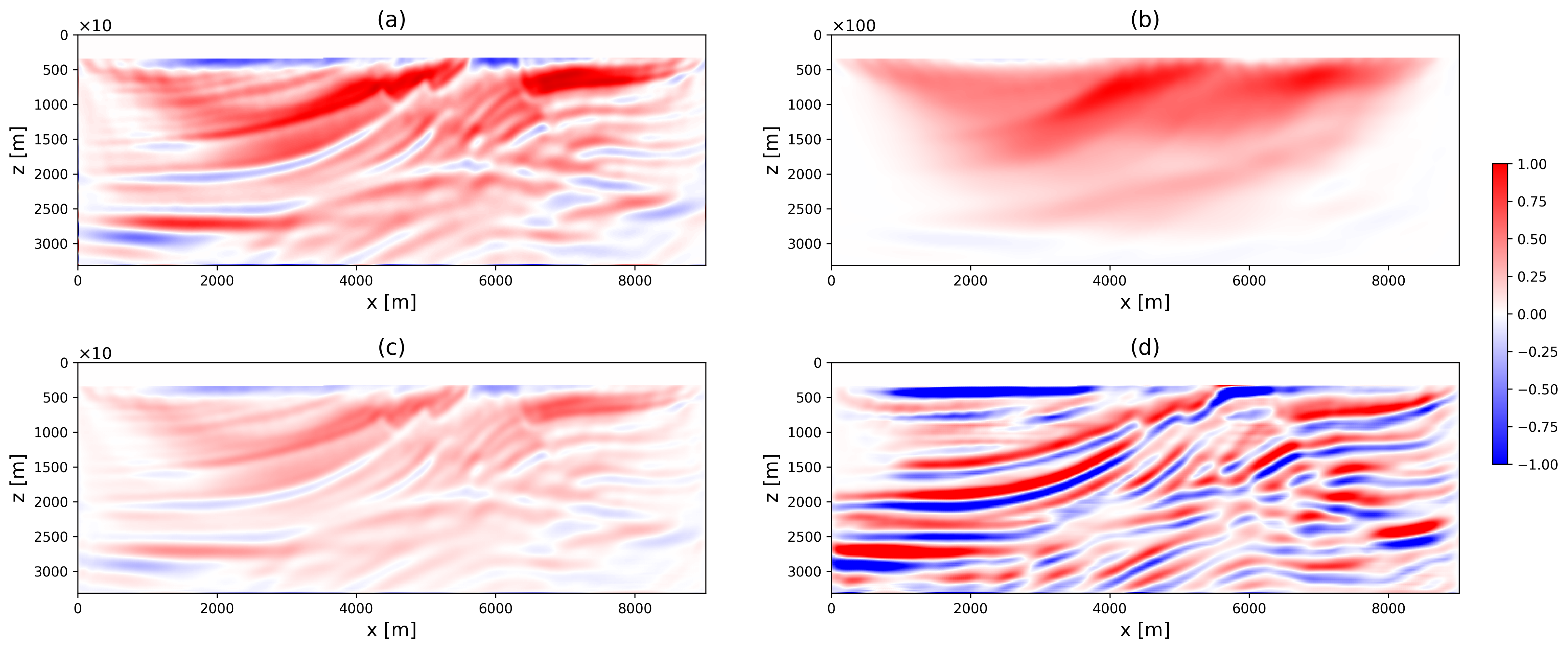} 
\caption{The components of our approach: a) the FWI gradient, b) $\delta m_1$, c) the predicted gradient by the network, and d) $\delta m$.} 
\label{fig_marm_iter_0} 
\end{figure}
To begin with, we consider the first iteration of the FWI process to detail the proposed approach. Starting from the FWI gradient (Figure~\ref{fig_marm_iter_0}a), we apply Born modeling and its adjoint to obtain the doubly-migrated image $\delta m_1$ in Figure~\ref{fig_marm_iter_0}b. Comparing the gradient to the doubly migrated image, we can see that the latter is of lower resolution and presents a stronger amplitude decay over depth; this is expected as the effect of geometrical spreading is accumulated multiple times as part of the Born modeling and its adjoint. Figure~\ref{fig_marm_iter_0}c shows the network output after training. This result highlights that the model has learned to map $\delta m_1$ to the gradient. Also, we show the normalized training loss for the first three iterations of FWI in Figure~\ref{fig_marm_train_loss} which highlights that the network is learning to map $\delta m_1$ to the gradient. Note that in the first FWI iteration the network is trained for 1000 iterations, but we only show the first 300 iterations just to make the plot more presentable. In addition, we should point out that the fitting of the gradient is not perfect (e.g., there is an amplitude loss) which could be a result of not having enough information to go from $\delta m_1$ to the gradient. Comparing the network's gradient to the FWI gradient we can say that the network has learned an approximation of the inverse Hessian. Thus, if we apply the network to the FWI gradient, we get a Hessian corrected update as shown in Figure~\ref{fig_marm_iter_0}d. For certain, we do not expect the network to learn the exact inverse Hessian due to the linear link between the gradient and $\delta m_1$ whilst the FWI inverse Hessian is non-linear. Not surprisingly, the update presents more balanced contributions throughout the entire depth axis and at the edges of the horizontal axis compared to the gradient; in other words, the effect of the acquisition geometry and geometrical spreading has been compensated for by the network.\\
To demonstrate the effectiveness of our approach, we compare its speed of convergence with that of a gradient-based method with Barzilai-Borwein (BB) update~\cite{10.1093/imanum/8.1.141} and a quasi-Newton method, namely L-BFGS~\cite{liu1989limited}. The BB method is a two-point step size gradient method used to approximate the quasi-Newton method. More precisely,  the Hessian is approximated with a scalar using the difference between two evaluations of the gradient and two solutions
\begin{equation}\label{eq:eq16}
\alpha_k^{long} = \frac{\Delta x \cdot \Delta x}{\Delta x \cdot \Delta g}
\end{equation}
\begin{equation}\label{eq:eq17}
\alpha_k^{short} = \frac{\Delta x \cdot \Delta g}{\Delta g \cdot \Delta g}
\end{equation}
where $\Delta x$ and $\Delta g$ are the differences between two solutions and two gradients, respectively. In our implementation we used $\alpha^{short}$. For nonlinear problems, the BB method tends to converge faster than the conjugate-gradient methods as demonstrated in~\cite{10.1007/0-387-24255-4_10}.\\
In order to provide a fair comparison in terms of computational cost for the different algorithms, the data misfit loss is plotted as function of the the number of wave equation solves (instead of the iteration itself). In our approach, in fact, every FWI iteration requires solving the wave equation twice more than in the conventional approaches. Alongside inspecting the data misfit loss, we also assess the quality of the updated model with respect to the true model (whenever possible) using the structural similarity index measure (SSIM)~\cite{1284395}; this choice is dictated by the fact that a small vertical (or horizontal) shift in the predicted model could lead to a small element-wise loss (e.g., mean squared error), even if the model closely resembles the true one.\\
As shown in Figure~\ref{fig_loss_marm} for the data loss and the SSIM index, the rate of convergence as function of the number of wave equations solved is higher for the proposed approach than for the BB and L-BFGS methods. Looking at the models after the first iteration of FWI, we observe that the predicted model from our method (Figure~\ref{fig_marm_iter1}b) is already of higher  quality than that obtained from the BB method (Figure~\ref{fig_marm_iter1}a). The shallow layers are better defined, and even the deeper parts of the model (at a depth of 3000 m or greater) are already updated. Moreover, looking at the final predicted models for the BB, L-BFGS, and $\delta m$ methods, respectively (Figure~\ref{fig_marm_iter_last}), it is clear that our method managed to better update the deeper part of the subsurface. This is expected when the Hessian is included in the model update as of the case with L-BFGS and $\delta m$, but $\delta m$ resolved the deeper part better when comparing the two approaches to the true model. Moreover, the update in the deeper part with $\delta m$ did not overestimate the width of the high velocity layers looking at the two ridges in the right and left of the deeper part of the model (depicted in dark red), for example.
\subsection{Volve synthetic}
\begin{figure}
\centering
\includegraphics[width=\linewidth]{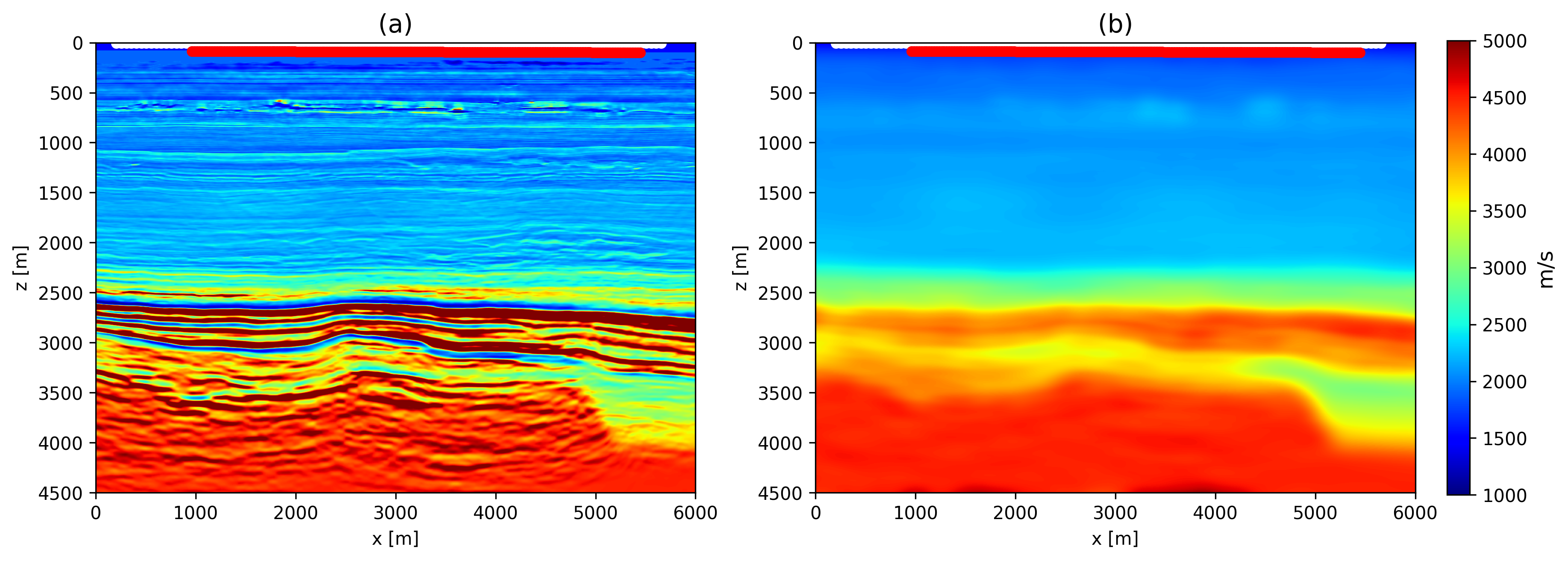} 
\caption{a) The true Volve synthetic model and b) its smoothed version. The arrays of sources and receivers are shown in white and red, respectively. } 
\label{fig_volve_sy} 
\end{figure}
\begin{figure}
\centering
\includegraphics[width=\linewidth]{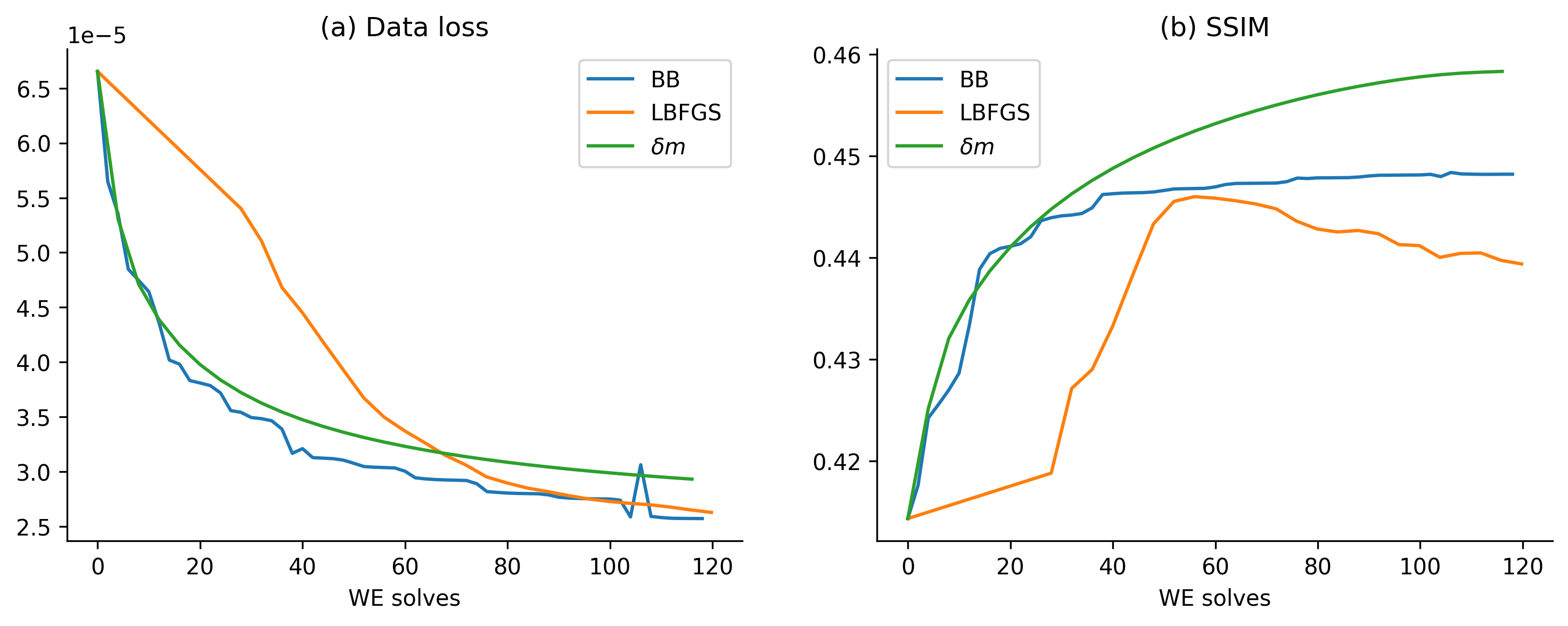} 
\caption{Data loss and SSIM index for FWI on the Volve synthetic dataset using the BB method, L-BFGS, and $\delta m$.} 
\label{fig_loss_volve_synth} 
\end{figure}
\begin{figure}
\centering
\includegraphics[width=\linewidth]{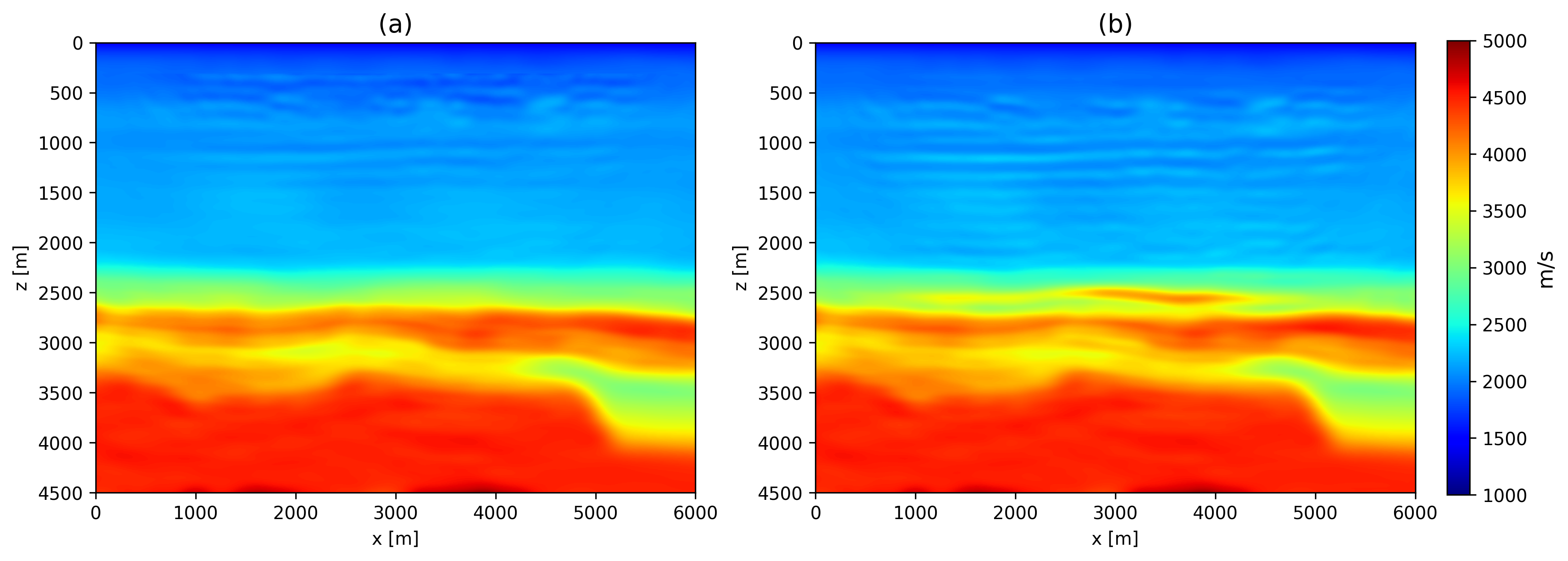} 
\caption{Third model update using a) the BB method, and b) $\delta m$ for Volve synthetic.} 
\label{fig_volve_synth_iter3} 
\end{figure}
\begin{figure*}
\centering
\includegraphics[width=\linewidth]{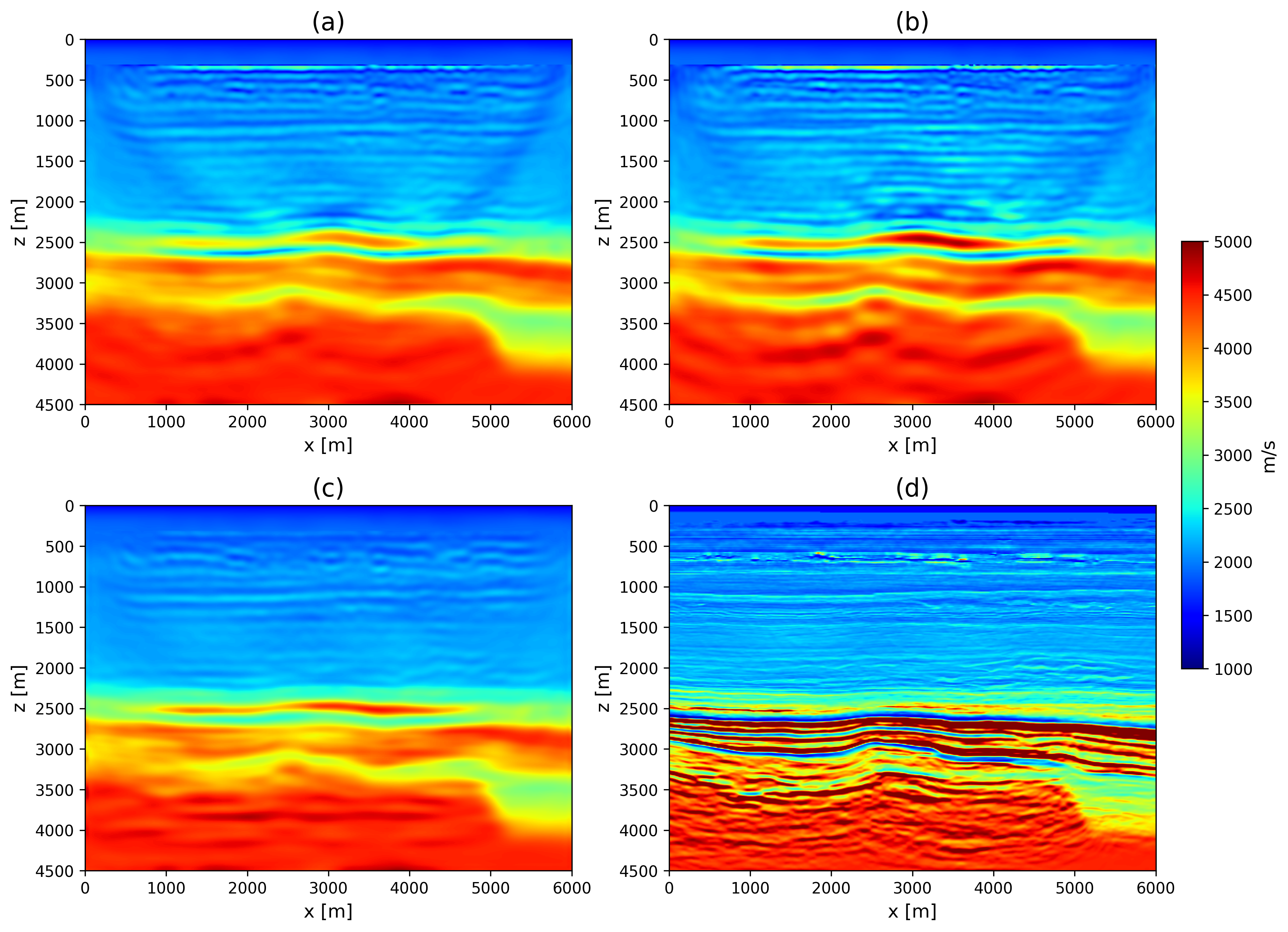} 
\caption{Final model update: a) using the BB method, b) using L-BFGS, and c) using $\delta m$.} 
\label{fig_volve_synth_iter_last} 
\end{figure*}
In our second example, we consider a synthetic dataset created to closely resemble the openly available Volve field data. This dataset has been created in~\cite{9785892} as follows: first, a high-resolution acoustic impedance model is obtained by applying post-stack seismic inversion to a 2D section of the migrated Volve field data. Next, the inverted acoustic impedance model is converted into a velocity model with the aid of well logs and used to simulate a synthetic dataset with a 20 Hz Ricker wavelet. An array of 110 sources, spaced 50 m apart and positioned at a depth of 6 m, along with an array of 180 receivers, spaced 25 m apart at depths ranging from 86 m to 99 m, are used to model the data.  Figure~\ref{fig_volve_sy} shows the inverted velocity model, which we refer to here as the true Volve synthetic velocity model and its smoothed version used as an initial model for FWI.

For the Volve synthetic, which is more challenging given its more realistic data and model, and larger model size, we utilized the data up to 7 Hz and we estimated the wavelet from it. Inspecting the data loss curves in Figure~\ref{fig_loss_volve_synth}a, our approach converges at a similar rate as the BB method; however the SSIM curves show that our method's estimated model is of a superior quality compared to that of the BB method (Figure~\ref{fig_loss_volve_synth}b); this is especially the case for later iterations where the BB model seems to be stagnant, whilst our model continues to improve. Even though L-BFGS reached a similar data residual to the BB and $\delta m$ methods at the end, the SSIM shows that the inverted model started to degrade half-way through the iterations. Also, looking at the updated models at the third (we chose the third iteration rather than the first because the difference is easier to spot in the third iteration.) and last iterations (Figures~\ref{fig_volve_synth_iter3} and~\ref{fig_volve_synth_iter_last}) using both approaches confirms that our approach resulted in a better inversion since in our approach the effect of the limited aperture is less pronounced (if any), and between 2200 to 2700 m depth the velocity estimation is better in general. In addition, our approach is not just a scalar approximation for the Hessian (like the BB step size) which is evident by the correction for limited aperture in the acquisition as shown in Figure~\ref{fig_volve_synth_iter_last}c. 
\subsection{Volve field data}
\begin{figure}
\centering
\includegraphics[width=\linewidth]{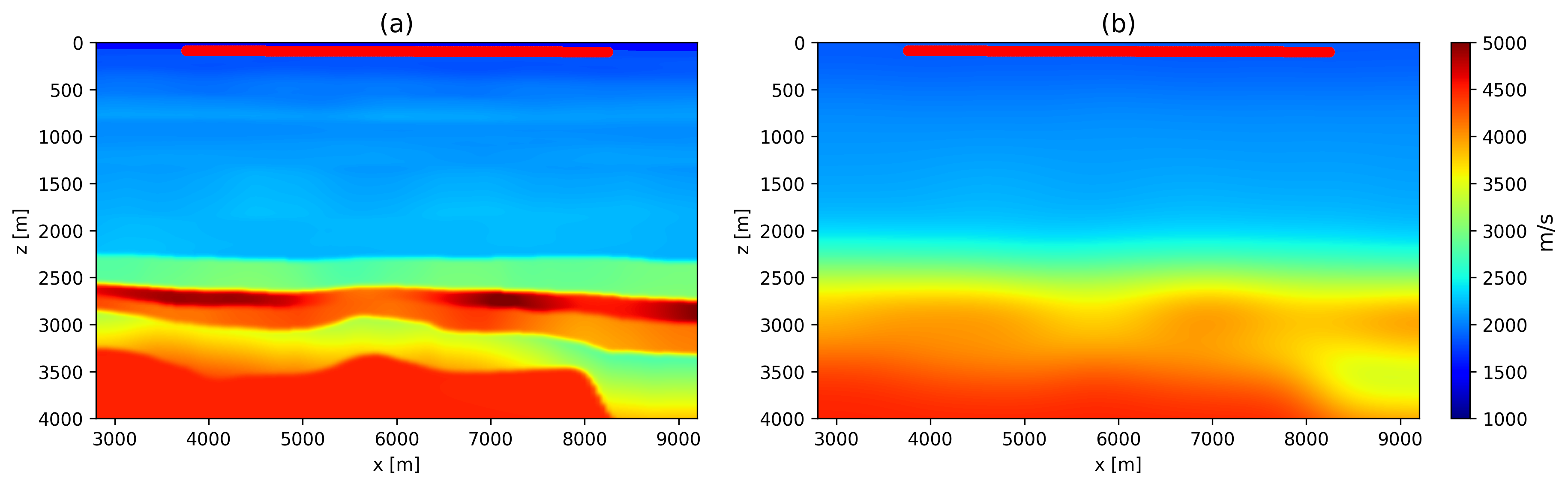} 
\caption{a) the velocity model obtained for Volve using tomography, and b) its smoothed version. The arrays of sources and receivers are shown in red.} 
\label{fig_vlove} 
\end{figure}
\begin{figure}
\centering
\includegraphics[width=\linewidth]{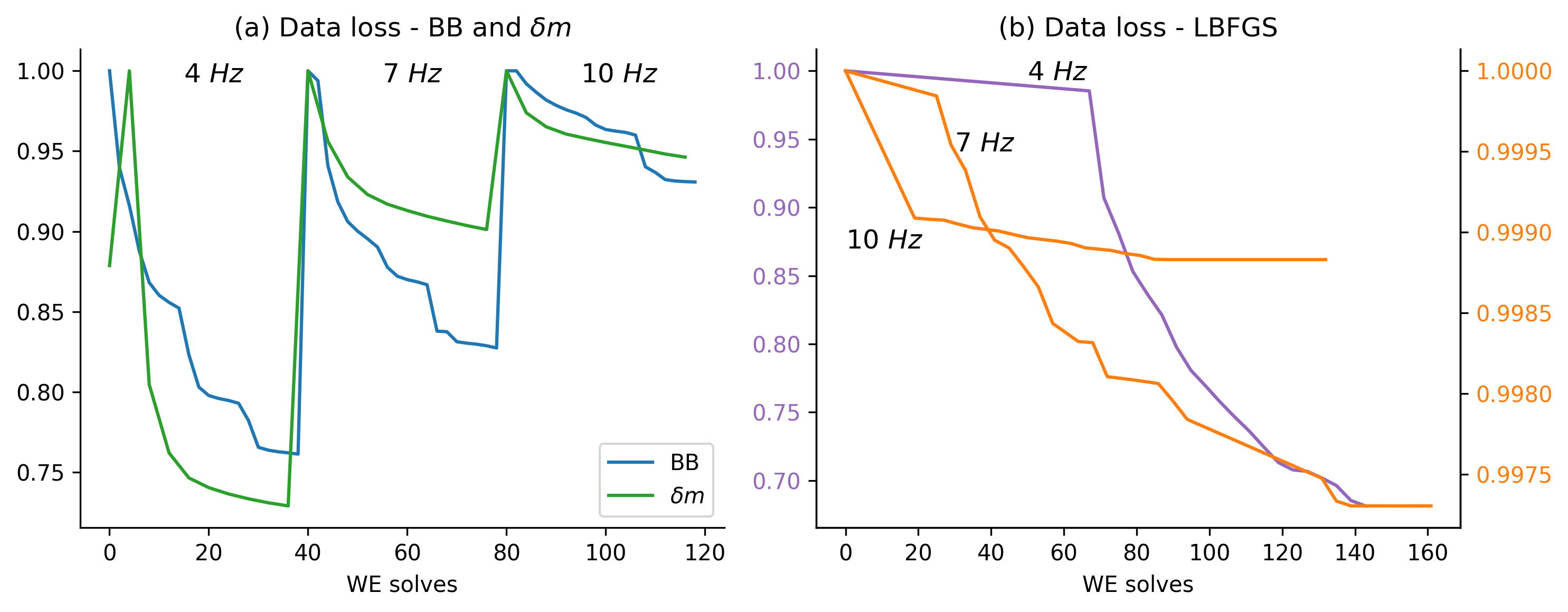} 
\caption{Data loss for FWI on the Volve dataset a) using the BB method and $\delta m$, and b) using L-BFGS.} 
\label{fig_loss_volve} 
\end{figure}
\begin{figure}
\centering
\includegraphics[width=\linewidth]{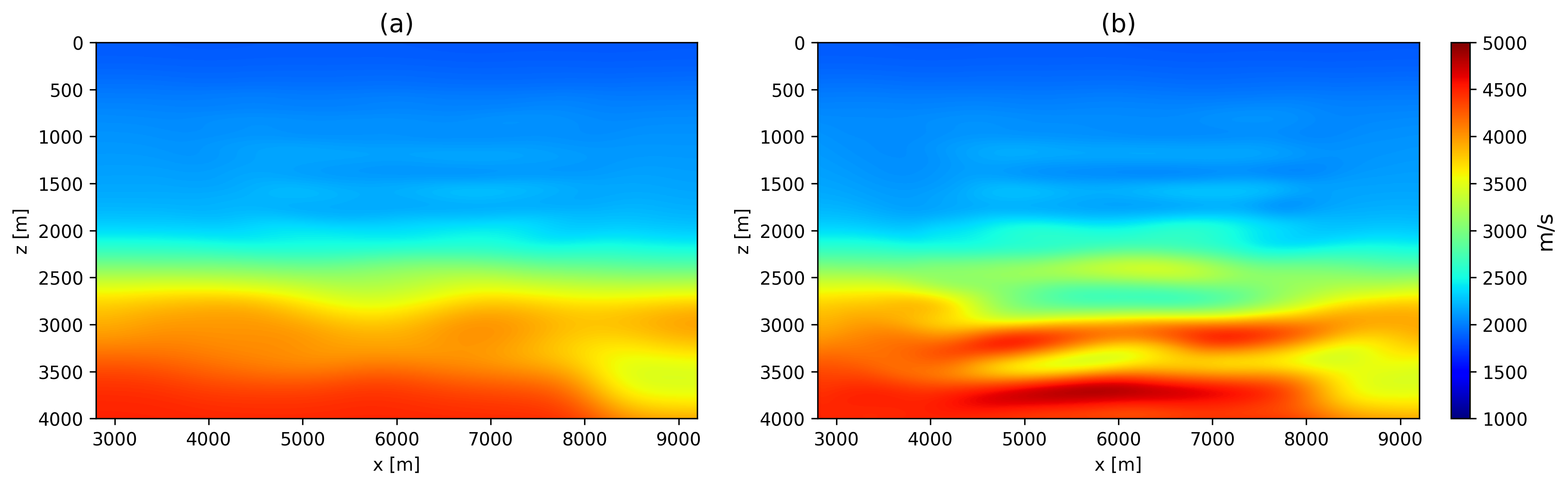} 
\caption{First model update: a) using the BB method, and b) using $\delta m$ for Volve.} 
\label{fig_volve_iter0} 
\end{figure}
\begin{figure*}
\centering
\includegraphics[width=\linewidth]{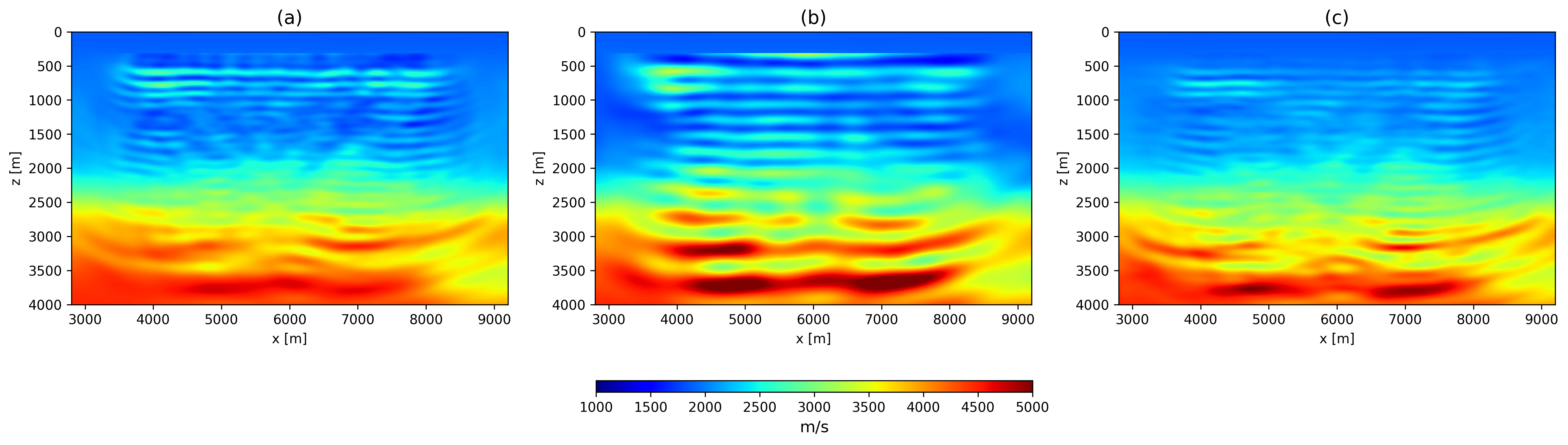} 
\caption{Final model update: a) using the BB method, b) using L-BFGS, and c) using $\delta m$ for Volve.} 
\label{fig_volve_iter_last} 
\end{figure*}
\begin{figure*}
\centering
\includegraphics[width=\linewidth]{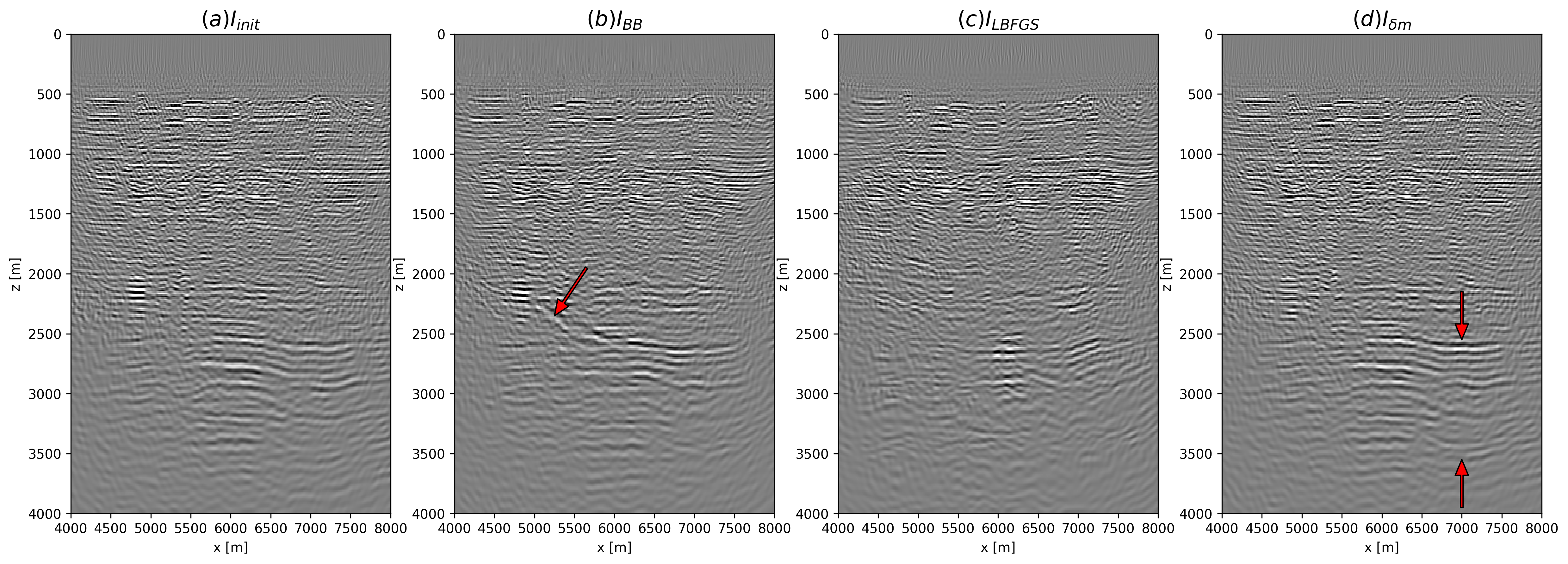} 
\caption{RTM images computed using: a) the initial model, b) the updated model with the BB method, c) the updated model with L-BFGS, and d) the updated model with $\delta m$ for Volve.} 
\label{fig_rtm_volve} 
\end{figure*}
\begin{figure}
\centering
\includegraphics[width=0.8\linewidth]{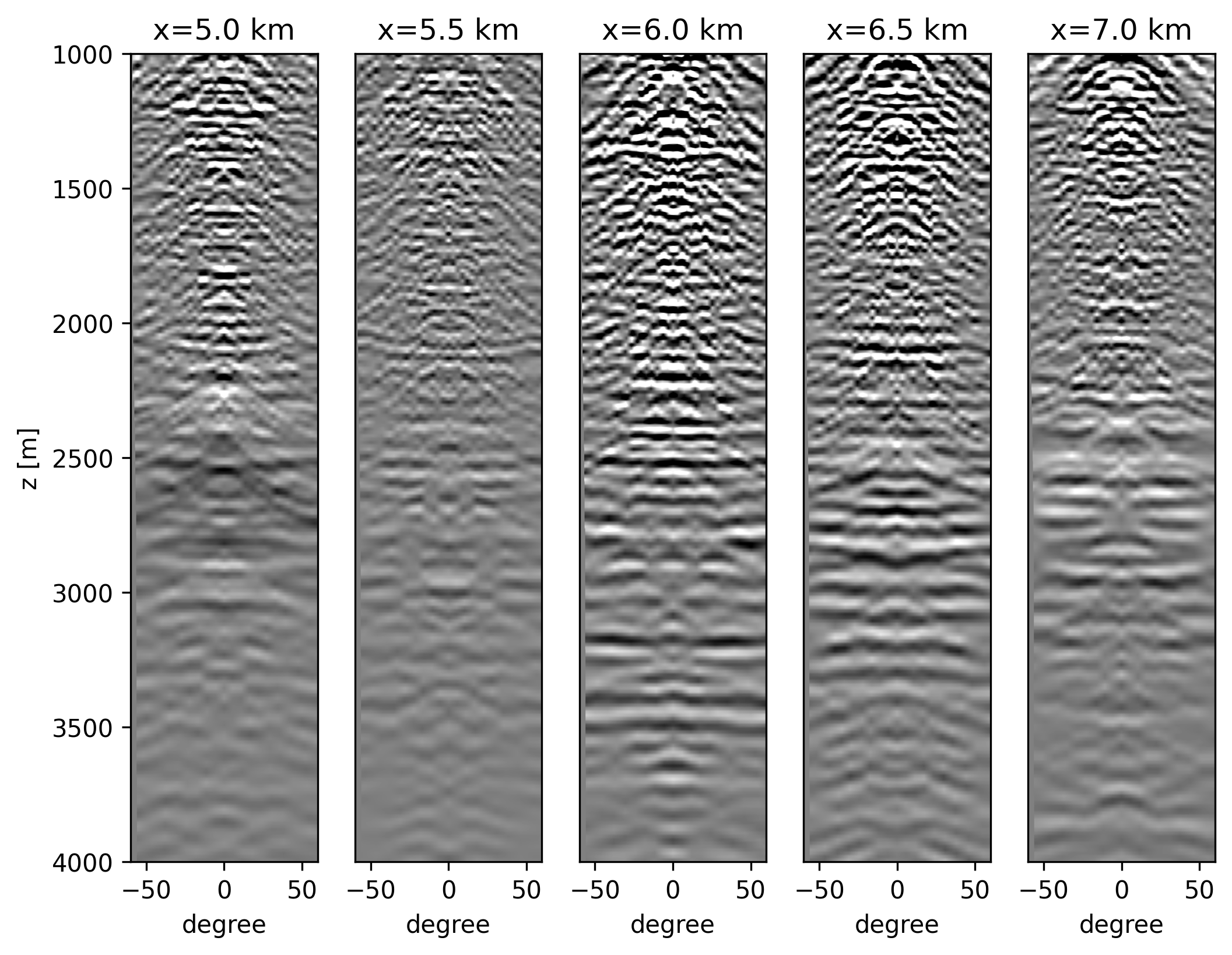} 
\caption{ADCIGs computed using the initial model at 5.0, 5.5, 6.0, 6.5, and 7.0 km in the $x$ direction for Volve.} 
\label{fig_adcig_init_volve} 
\end{figure}
\begin{figure}
\centering
\includegraphics[width=0.8\linewidth]{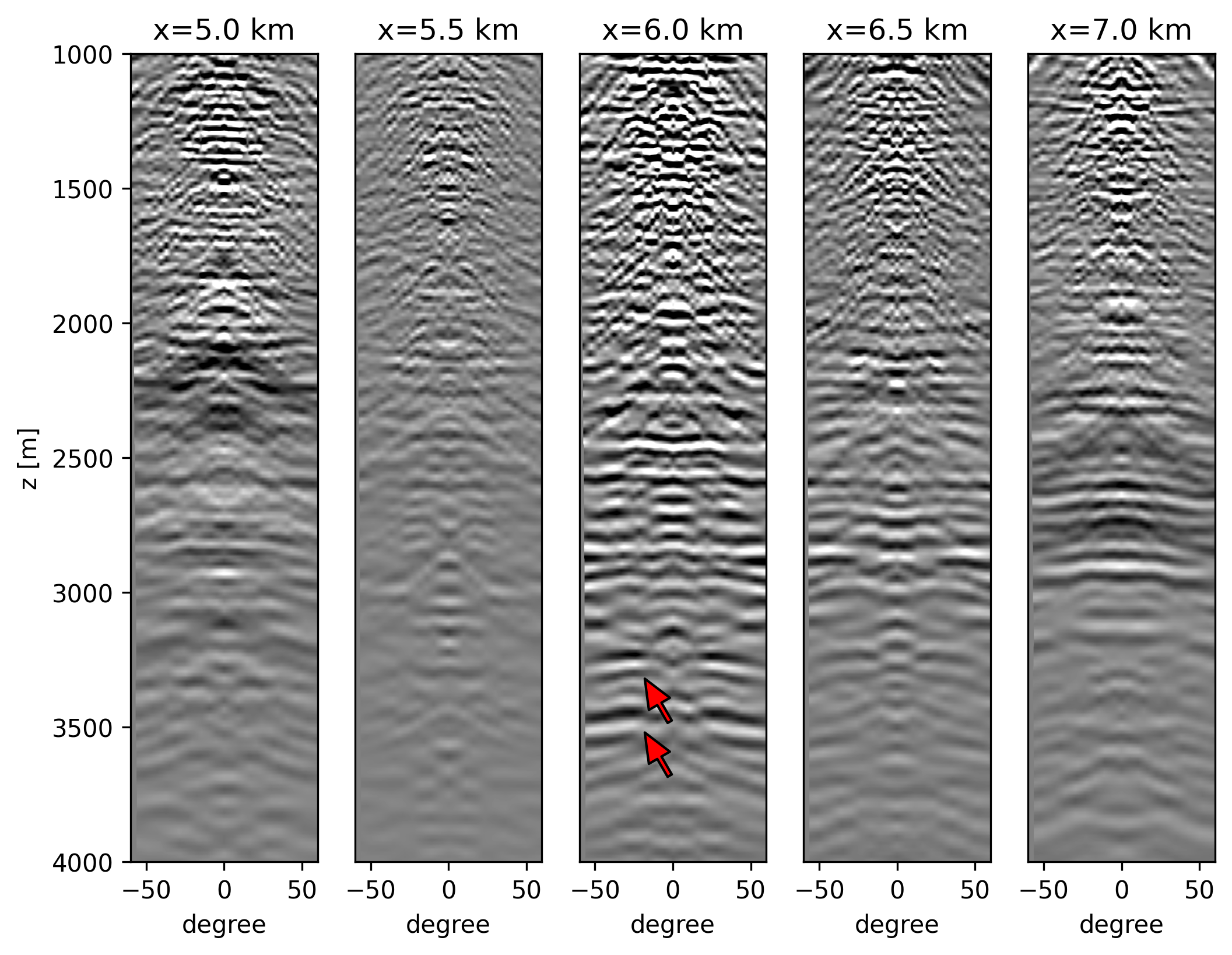} 
\caption{ADCIGs computed using the updated model with the BB method at 5.0, 5.5, 6.0, 6.5, and 7.0 km in the $x$ directionfor Volve.} 
\label{fig_adcig_bb_volve} 
\end{figure}
\begin{figure}
\centering
\includegraphics[width=0.8\linewidth]{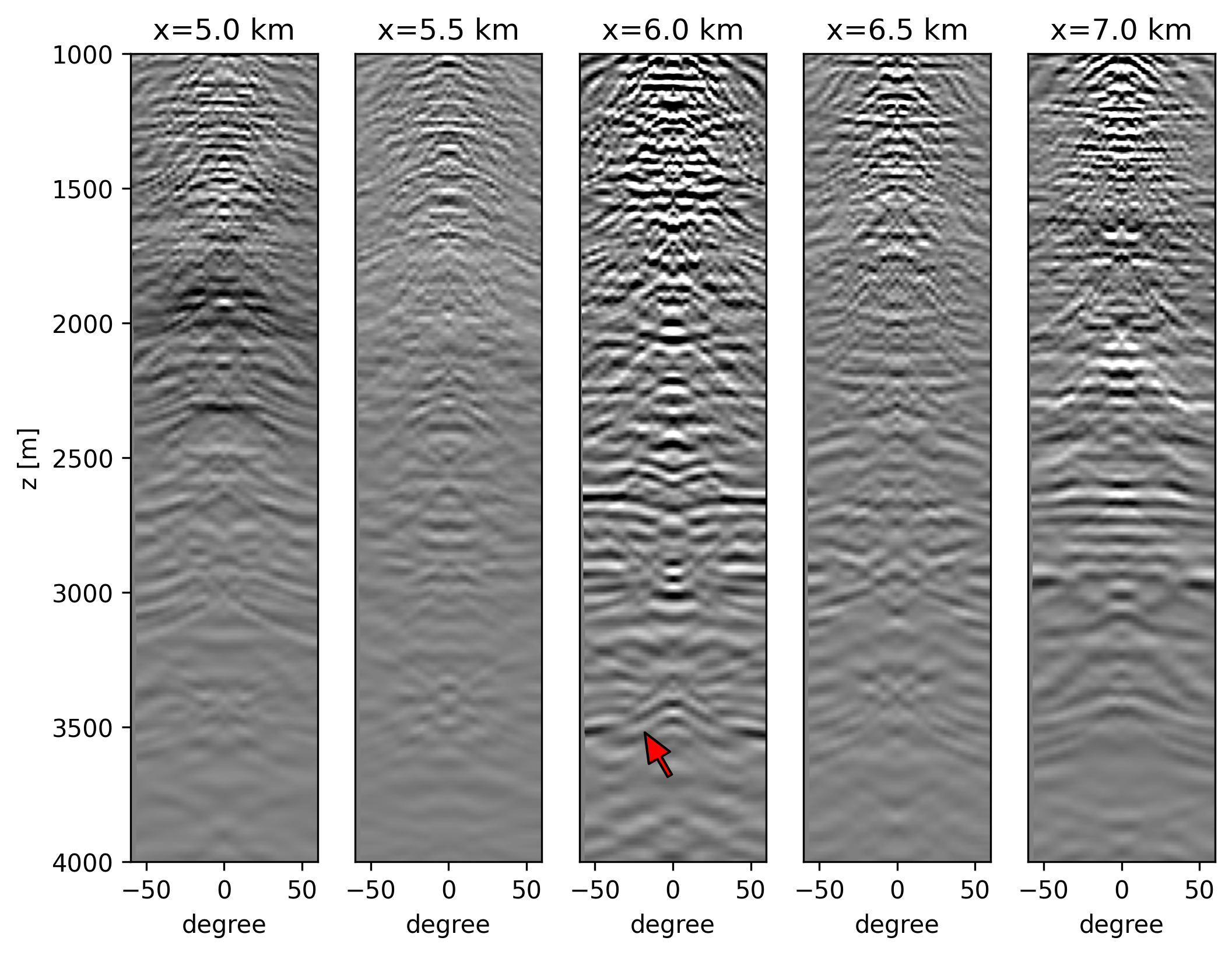} 
\caption{ADCIGs computed using the updated model with L-BFGS at 5.0, 5.5, 6.0, 6.5, and 7.0 km in the $x$ direction for Volve.} 
\label{fig_adcig_lbfgs_volve} 
\end{figure}
\begin{figure}
\centering
\includegraphics[width=0.8\linewidth]{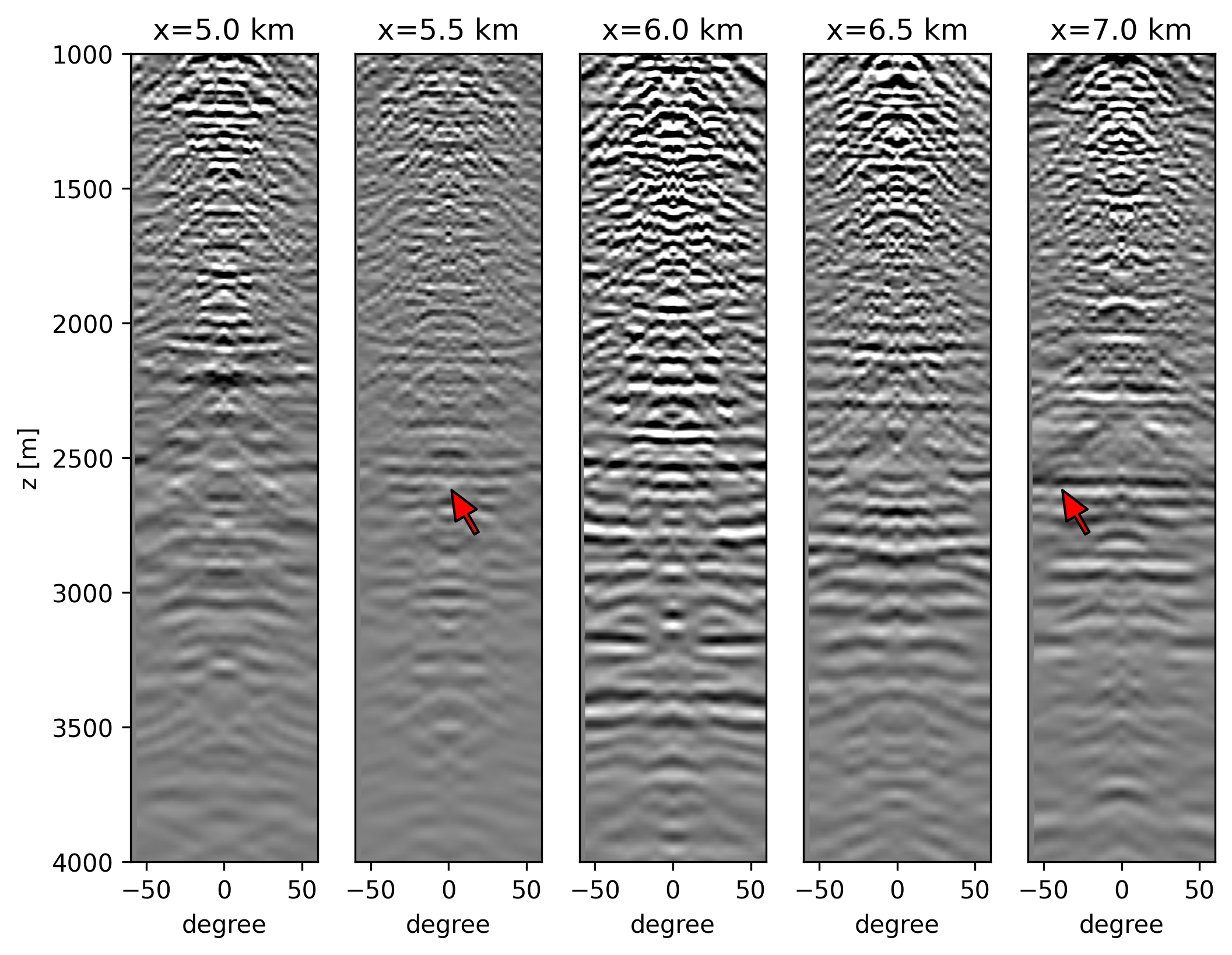} 
\caption{ADCIGs computed using the updated model with $\delta m$ at 5.0, 5.5, 6.0, 6.5, and 7.0 km in the $x$ direction for Volve.} 
\label{fig_adcig_dm_volve} 
\end{figure}
\begin{figure*}
\centering
\includegraphics[width=\linewidth]{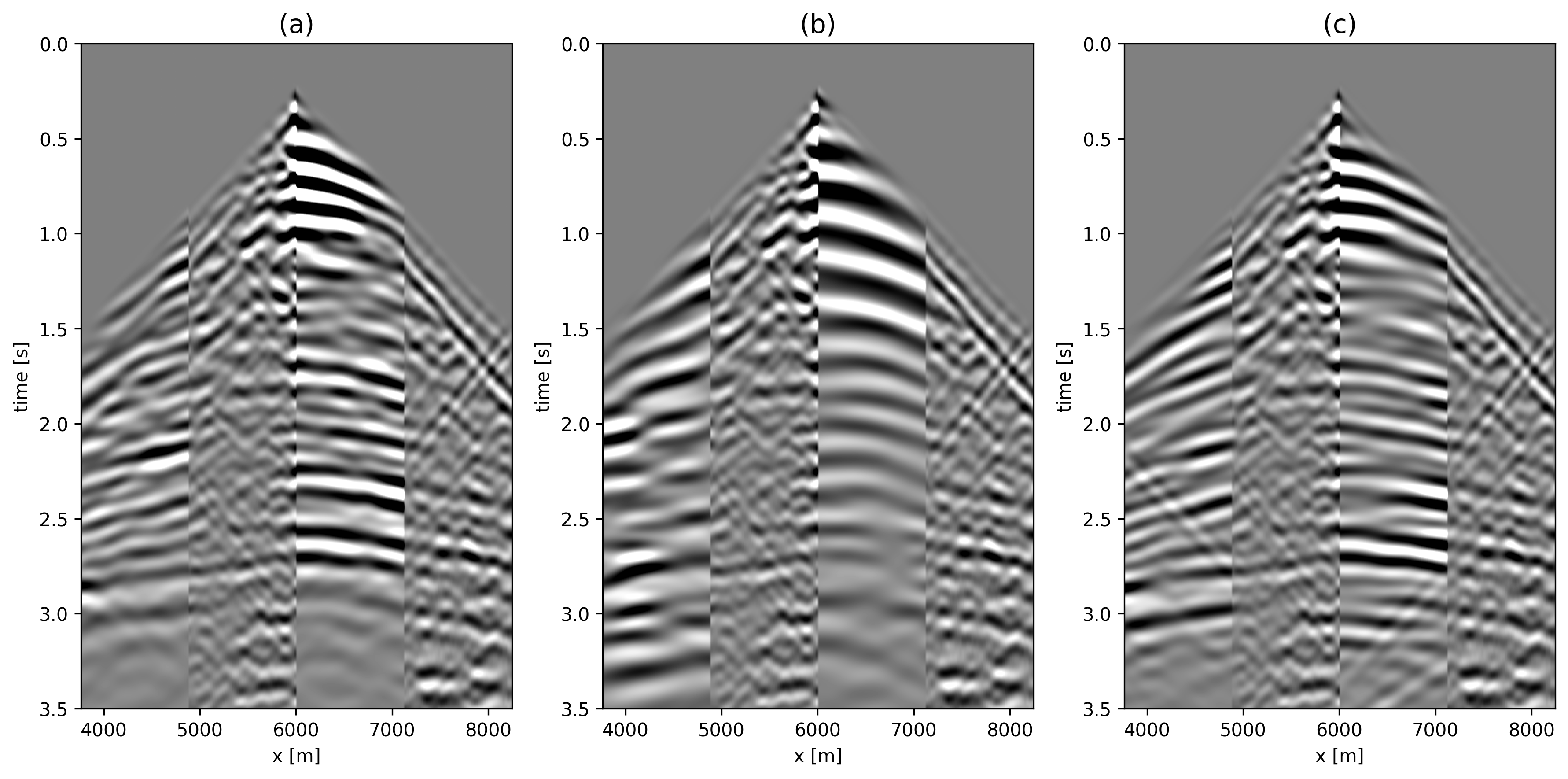} 
\caption{Observed data compared to simulated data computed with the model inverted a) using the BB method, b) using L-BFGS, and c) using $\delta m$. First and third quarters of each panel are the simulated data while the second and fourth quarters are the observed data. } 
\label{fig_data_volve} 
\end{figure*}
In 2008, the production of oil commenced in the Volve oil field located in the Norwegian North Sea and continued until its shutdown in 2016. From the 3D OBC survey acquired by Statoil in 2010 for Volve, we selected a 2D line to test our algorithm. Since we removed multiples by multi-dimensional deconvolution (MDD), the processed data has now co-located sources and receivers at the seafloor. Thus, this 2D line now contains 180 sources and receivers, positioned at depths ranging from 86.8 m to 99.2 m, with a spacing of 25 m. In addition, we applied other processing methods such as denoising, vector-fidelity corrections, anti-aliasing filtering, and a scaling by $\sqrt{t}$ to manage the conversion from 3D to 2D due to geometrical spreading. For a detailed description of the processing sequence, the reader is referred to~\cite{eage:/content/papers/10.3997/2214-4609.201413355,10.1093/gji/ggv528}. Figure~\ref{fig_vlove} shows a 2D section of the tomographic velocity model provided as part of the open Volve dataset along the line of sources and receivers selected for this study in red, and its smoothed version used as starting model for FWI.

Working with field data is more challenging due to factors such as unknown wavelet, insufficient frequency content in the data, different physics from the one used in the numerical modeling which leads to events in the data that cannot be matched/reproduced, etc. Thus, after pre-processing the data to be suitable for seismic processing and imaging as described earlier, we approximated the shape and scale of the wavelet from the data and ran multi-scale FWI for 4, 7, and 10 Hz maximum frequency as commonly done in practical applications of FWI. Figure~\ref{fig_loss_volve}a shows the normalized data loss for the three frequencies from lowest to highest for the BB method and $\delta m$ while Figure~\ref{fig_loss_volve}b is for L-BFGS. Since we are plotting three different runs of FWI and we have less control on the number of iteration when running the L-BFGS optimization, we plotted the curves for L-BFGS in a separate plot. For 4Hz and 10 Hz our approach converges faster than the BB method, whilst for 7Hz it seems to be slightly slower. In contrast, L-BFGS takes more time to catch up with the other two methods for 4 Hz and does not perform as well in the other two frequency scales. Note that large increases in data loss when the frequency range is changed are expected because there is a jump in the frequency content of the data which is not represented in the updated velocity model. Figure~\ref{fig_volve_iter0} shows the first updated model for the BB and $\delta m$ methods: it is clear that our approach did not only provide a better update for the shallow part of the model, but it also was able to update the deep layers. Similarly, when looking at the final updated models in Figure~\ref{fig_volve_iter_last} for the three approaches, we can conclude that our approach has led to more significant changes in the deeper layers, whilst at the same time the velocities of the shallower layers are not overestimated as in the BB method and L-BFGS.

To validate our observations, we finally performed imaging using both the initial and the three estimated velocity models (Figure~\ref{fig_rtm_volve}). The image produced using the velocity model from the BB method presents an artifact between around $2-2.5$ km depth (red arrow in Figure~\ref{fig_rtm_volve}b), which does not appear in the other three images. Also, the image produced from L-BFGS (Figure~\ref{fig_rtm_volve}c) is of lower quality as the reflectors in the deeper part are less focused. Most likely, this is because L-BFGS was more aggressive in its velocity update. Lastly, comparing the image produced using our velocity model with the one obtained from the initial velocity model, we notice that the former presents more continuous reflectors (red arrow in Figure~\ref{fig_rtm_volve}d); moreover, the reflector at around $2.5$ km has been slightly shifted downward. In addition, Figures~\ref{fig_adcig_init_volve},~\ref{fig_adcig_bb_volve},~\ref{fig_adcig_lbfgs_volve}, and~\ref{fig_adcig_dm_volve} show the angle-domain common image gathers (ADCIGs) produced using the same three velocity models used for imaging. The image gathers produced from the initial model are in general flat; this suggests that the initial model is already good in terms of kinematic which is because the tomographic model is highly accurate~\cite{eage:/content/papers/10.3997/2214-4609.201402177}, but we smoothed it to mimic the scenario where such a model is not available. However, the angle gathers for the BB method are curving down as pointed out by the red arrows in Figure~\ref{fig_adcig_bb_volve} (i.e., the updated velocity model is not better). On the other hand, the angle gathers produced with the velocity model estimated by our approach show similar flatness as the ones from the initial model but with more focused reflectors and better continuity, indicated by red arrows in Figure~\ref{fig_adcig_dm_volve}. As for L-BFGS, like the RTM image, the gathers are not focused and not flat, see the red arrow in Figure~\ref{fig_adcig_lbfgs_volve} for example. Finally, Figures~\ref{fig_data_volve}a-c compares the observed data with the simulated data using the inverted models with the BB methos, L-BFGS, and $\delta m$, respectively. In each panel, the first and third quarters are the simulated data, and the second and the fourth quarters are the observed data. In general, the BB method, and $\delta m$ data align more accurately and resemble better the field data than L-BFGS data. However, $\delta m$ data has stronger events in the deeper part.


\section{Discussion}
The novelty of this work stems from the fact that we embed the concept of approximating the inverse of the Hessian into a neural network, which is trained to map a Hessian-adjusted gradient to the FWI gradient. The effect of the Gauss-Newton Hessian is added to the FWI gradient by means of Born modeling and its adjoint, and the role of the network is that of approximating the inverse of the Hessian, with the goal of producing an improved update vector at each iteration of FWI. Since we expect the inverse of the Hessian to slowly change from iteration to iteration, we leverage the ability of neural networks to learn a task and adapt quickly to a similar, but slightly different task via transfer learning. This would not be the case for standard approaches with matching filters, as they would have to be performed from scratch at every FWI iteration. Furthermore, since the inverse of the Hessian can be considered as a deblurring operator that can increase the resolution of images, the use of neural networks to approximate it is attractive due to their success in super-resolution applications.

Despite the aforementioned advantages, our method comes also with some shortcomings. First, the Hessian that links the gradient with the doubly migrated image based on Born modeling and its adjoint is an approximation of the FWI Hessian. Thus, the resulting approximation of the inverse of the Hessian most likely does not provide as fast convergence (which is, also, the case with most matching filters methods) as the actual inverse of the FWI Hessian, even though we expect this to be better than having no Hessian at all. Second, the bandwidth of the signal we are attempting to recover (the full FWI update) is larger than what the network is provided within training. That is due to the fact that the FWI gradient and the doubly migrated image are 'convolved` by one and two wavelets, respectively. Thus, the doubly migrated image has the same frequency support of the FWI gradient or even less, and both have less support than the sought after scattering potential. Thus, even though the network has learned to produce an enhanced version of the gradient, there is no guarantee that this can extend the frequency content beyond that present in the training data. Third, our approach presents a computational bottleneck as it requires to perform an additional step of modeling-migration at each FWI iteration, as well as training or fine-tuning the network. Applying the network to the gradient once it is trained is fast, however the training cost is not negligible.

We have applied our method to three examples, two with synthetic datasets and one with the Volve field dataset. In general, our approach converges faster (or has a similar convergence rate) than the quasi-Newton methods (BB and L-BFGS). However, as discussed in the two Volve examples, our approach resulted in a better inverted model as assessed by the SSIM index in the synthetic case or by the different imaging products in the field case. Field data are contaminated by noise and unwanted events, and thus, we do not wish to perfectly fit the data. Thus, having a neural network embedded within the FWI iteration can have a regularization effect, and thus, we might avoid fitting the noise in the data as supported by our Volve example.

Finally, we have observed experimentally that our approach works best when inverting low frequency data; however, when we start FWI from higher frequencies, our method does not perform as well as when starting from low frequencies. Nevertheless, our method is still applicable to the multi-scale FWI, provided that we train the network from scratch at the start of each new frequency band. On the other hand, even though our approximation of the Hessian entails a cost equals to one conventional FWI iteration to obtain the remigrated image, we have a better model update from the start as illustrated in the previous examples. Thus, we can use our approximation of the Hessian in the early stage of FWI to steer the optimization problem into a better direction, and then,  continue with the conventional FWI. This approach will not only reduce the cost, but it is also more practical taking into account that the role of inverse of the Hessian is generally more important in the early iterations of FWI.


\section{Conclusions}
We have proposed a deep learning-based approach to approximate the inverse of the Hessian in the context of FWI. Our method is based on obtaining a remigrated image that is related to the gradient through the inverse of the Hessian by means of Born modeling and its adjoint; subsequently, a neural network is trained to learn the mapping from the remigrated image to the gradient; the trained network is then applied to the gradient to enhance its resolution and compensate for illumination effects. Numerical results on three different datasets show that the enhanced gradients can lead to an FWI algorithm with faster convergence and improved inverted models.


\section*{Acknowledgments}
The author thanks KAUST and the DeepWave Consortium sponsors for supporting this research, as well as Equinor and the Volve license partners for releasing the field dataset.




\bibliographystyle{unsrtnat}
\bibliography{References}
\nocite{ravasi2020pylops, richardson_alan_2023}


\section*{\textsc{Appendix: Network architecture}}
\begin{table}[!ht]
\caption{Summary of the network architecture}
\centering
\begin{tabular}{@{}lll}
\textbf{Layer} & \textbf{Operation} & \textbf{Channels} \\ \hline
Input Conv     & 2 $\times$ Conv2d + LeakyReLU & 128 \\ 
Downsampling 1 & AvgPool2d + 2 $\times$ Conv2d + LeakyReLU & 256 \\
Downsampling 2 & AvgPool2d + 2 $\times$ Conv2d + LeakyReLU & 512 \\
Downsampling 3 & AvgPool2d + 2 $\times$ Conv2d + LeakyReLU & 1024 \\
Downsampling 4 & AvgPool2d + 2 $\times$ Conv2d + LeakyReLU & 2048 \\ 
Upsampling 1   & ConvTranspose2d + 2 $\times$ Conv2d + LeakyReLU & 1024 \\
Upsampling 2   & ConvTranspose2d + 2 $\times$ Conv2d + LeakyReLU & 512 \\
Upsampling 3   & ConvTranspose2d + 2 $\times$ Conv2d + LeakyReLU & 256 \\
Upsampling 4   & ConvTranspose2d + 2 $\times$ Conv2d + LeakyReLU & 128 \\ 
Output Conv    & Conv2d & 1 \\ 
\hline
\end{tabular}
\label{tab:unet}
\end{table}








\end{document}